\documentclass[10pt,journal,english,twocolumn]{IEEEtran} 

\pdfoutput=1

\usepackage[T1]{fontenc}

\usepackage{babel}
\usepackage[babel]{microtype}
\usepackage[babel]{csquotes}
\usepackage{doi}

\usepackage[svgnames]{xcolor}

\usepackage{algorithm}
\usepackage{mathtools}
\usepackage{algpseudocode}
\usepackage{amsfonts}
\usepackage{tabularx}
\usepackage{tkz-euclide}
\usepackage{amsthm}

\usepackage{subfigure}
\usepackage{hhline}
\usepackage{booktabs}
\usepackage{pgfplots}
\usepackage{algpseudocode}
\usepackage[nointegrals]{wasysym}

\usepackage{siunitx}
\DeclareSIUnit{\dBm}{dBm}
\usepackage{xparse}

\usepackage{tikz}

\usetikzlibrary{shapes.geometric, arrows,calc,3d,spy}
\tikzstyle{box} = [rectangle, rounded corners, minimum width=2cm, minimum height=0.8cm,text centered, draw=black,]
\pgfplotsset{compat=newest}
\tikzstyle{box} = [rectangle, rounded corners, minimum width=2cm, minimum height=0.8cm,text centered, draw=black,]
\tikzstyle{label} = [minimum height=0.8cm,text centered, draw=none]

\tikzset{
>=stealth',
help lines/.style={dashed, thick},
axis/.style={<->},
important line/.style={ultra thick, smooth},
connection/.style={thick, dotted},
}

\tikzstyle{feature} = [rectangle, minimum width=2cm, minimum height=.6cm, align=center, text centered, draw=black, ]
\tikzstyle{layer} = [rectangle, rounded corners, minimum width=2.5cm, minimum height=.6cm, align=center, text centered, draw=black]
\usetikzlibrary{shapes.arrows}

\makeatletter 
\tikzoption{canvas is xy plane at z}[]{%
  \def\tikz@plane@origin{\pgfpointxyz{0}{0}{#1}}%
  \def\tikz@plane@x{\pgfpointxyz{1}{0}{#1}}%
  \def\tikz@plane@y{\pgfpointxyz{0}{1}{#1}}%
  \tikz@canvas@is@plane
}
\makeatother
\NewDocumentCommand{\DrawCubes}{O {} m m m m m m}{%
    \def\XGridMin{#2}
    \def\XGridMax{#3}
    \def\YGridMin{#4}
    \def\YGridMax{#5}
    \def\ZGridMin{#6}
    \def\ZGridMax{#7}
    \begin{scope}[canvas is xy plane at z=\ZGridMax]
      \draw [#1] (\XGridMin,\YGridMin) grid (\XGridMax,\YGridMax);
    \end{scope}
    \begin{scope}[canvas is yz plane at x=\XGridMax]
      \draw [#1] (\YGridMin,\ZGridMin) grid (\YGridMax,\ZGridMax);
    \end{scope}
    \begin{scope}[canvas is xz plane at y=\YGridMax]
      \draw [#1] (\XGridMin,\ZGridMin) grid (\XGridMax,\ZGridMax);
    \end{scope}
}%

\theoremstyle{plain}
\newtheorem{theorem}{Theorem}
\newtheorem*{theorem*}{Theorem}

\theoremstyle{remark}
\newtheorem{remark}{Remark}

\newtheoremstyle{example}{\topsep}{\topsep}{}{}{\itshape}{.}{ }{}
\theoremstyle{example}

\newtheorem*{example*}{Example}

\usepackage[backend=biber, style=ieee, url=false, isbn=false, dashed=false]{biblatex}
\bibliography{modernize-ref.bib}

\usepackage{xspace}
\newcommand{\InterferenceNet}{SINRnet\xspace}

\usepackage{hyperref}
\hypersetup{
	pdfauthor={Bile Peng, Karl-Ludwig Besser, Ramprasad Raghunath, Eduard A. Jorswieck},
	pdftitle={Approaching Globally Optimal Energy Efficiency in Interference Networks via Machine Learning},
	bookmarksnumbered=true,
	colorlinks=true,
	allcolors=.,
	urlcolor=MidnightBlue,
}

\usepackage[acronyms,nomain,xindy,nowarn]{glossaries}
\makeglossaries
\newacronym{bs}{BS}{base station}
\newacronym{ee}{EE}{energy efficiency}
\newacronym{wmmse}{WMMSE}{weighted minimum mean square error}
\newacronym{nn}{NN}{neural network}
\newacronym{sinr}{SINR}{signal-to-interference-noise ratio}
\newacronym{mimo}{MIMO}{multiple-input-multiple-output}
\newacronym{ofdma}{OFDMA}{orthogonal frequency-division multiple access}
\newacronym{ann}{ANN}{artificial neural network}
\newacronym{cn}{CN}{cellular network}
\newacronym{gnn}{GNN}{graphical neural network}
\newacronym{em}{EM}{expectation-maximization}
\newacronym{mse}{MSE}{mean squared error}
\newacronym{iid}{i.i.d.}{independent identically distributed}
\newacronym{sga}{SGA}{stochastic gradient ascent}
\newacronym{ngmn}{NGMN}{next generation mobile networks}
\newacronym{prb}{PRB}{physical resource block}
\newacronym{dnn}{DNN}{deep neural network}
\newacronym{sca}{SCA}{successive convex approximation}
\newacronym{csi}{CSI}{channel state information}
\newacronym{mrc}{MRC}{maximum ratio combining}
\newacronym{sic}{SIC}{successive interference cancellation}
\setacronymstyle{long-short}
\glsdisablehyper

\let\vec\mathbf
\let\mat\mathbf
\newcommand*{\reals}{\ensuremath{\mathbb{R}}}
\DeclareMathOperator{\relu}{ReLU}

\definecolor{change}{HTML}{0096b8}
\newenvironment{change}{\color{black}}{}

\title{Approaching Globally Optimal Energy Efficiency in Interference Networks via Machine Learning}

\author{Bile Peng, \IEEEmembership{Member,~IEEE}, Karl-Ludwig Besser, \IEEEmembership{Member,~IEEE}, Ramprasad Raghunath, \IEEEmembership{Student Member,~IEEE}, and Eduard A. Jorswieck, \IEEEmembership{Fellow,~IEEE}%
\thanks{B. Peng, R. Raghunath and E. A. Jorswieck are with the Department of Information Theory and Communication Systems, TU Braunschweig, Schleinitzstr. 22, 38112 Braunschweig, Germany (e-mail: \{b.peng, r.raghunath, e.jorswieck\}@tu-braunschweig.de).
K.-L. Besser was with the Institute for Communications Technology, Technische Universität Braunschweig, 38106 Braunschweig, Germany, and is now with the Department of Electrical and Computer Engineering, Princeton University,  Princeton, NJ 08544, USA (email: {kb1717}@princeton.edu).}%
\thanks{The work is supported by the Federal Ministry of Education and Research Germany (BMBF) as part of the 6G Research and Innovation Cluster 6G-RIC under Grant 16KISK031.}%
\thanks{This paper is published on IEEE Transactions on Wireless Communications (DOI: 10.1109/TWC.2023.3269770).}
}

\begin{document}
\maketitle

\begin{abstract}
This work presents a machine learning approach to optimize the \gls{ee} in a multi-cell wireless network.
This optimization problem is non-convex and its global optimum is difficult to find.
In the literature, either simple but suboptimal approaches or optimal methods with high complexity are proposed.
In contrast, we propose an unsupervised machine learning framework to approach the global optimum.
While the \gls{nn} training takes moderate time,
application with the trained model requires very low computational complexity.
In particular, we introduce a novel objective function based on stochastic actions to solve the non-convex optimization problem.
Besides, we design a dedicated \gls{nn} architecture \InterferenceNet for the power allocation problems in the interference channel that is permutation-equivariant.
We encode our domain knowledge into the \gls{nn} design and shed light into the black box of machine learning.
Training and testing results show that the proposed method without supervision and with reasonable computational effort achieves an \gls{ee} close to the global optimum found by the branch-and-bound algorithm
and outperform the \gls{sca} algorithm.
Hence, the proposed approach balances between computational complexity and performance.
\end{abstract}

\begin{IEEEkeywords}
Energy efficiency, non-convex optimization, machine learning, permutation-equivariance, reparameterization trick.
\end{IEEEkeywords}

\glsresetall

%

\section{Introduction}
\label{sec:intro}

Transmit power control belongs to the most fundamental problems in research on cellular mobile networks~\cite{chiang2008power} and it is an important measure to optimize objectives such as weighted sum-rate~\cite{zhang2011weighted}, \gls{ee}~\cite{matthiesen2020globally} and fairness~\cite{ghazanfari2020enhanced}.
These problems share some significant similarities, among which the non-convexity due to the fractional nature of the \gls{sinr} expression is a major difficulty, because multiple local optima could exist and convex optimization techniques are not applicable.

\begin{change}
This paper focuses on the \gls{ee} optimization since the energy consumption is a key performance indicator of green communication in the fifth-generation (5G) communication and beyond.
According to the \gls{ngmn} alliance 5G white paper~\cite{alliance20155g},
the \gls{ee} is required to be improved by a factor of 2000 compared to current wireless networks.
\end{change}
Among different techniques, transmit power control is an important approach to improve the \gls{ee} in the air interface.
On the other hand, given the similarities of the above-mentioned power control problems in cellular networks and the fact that the \gls{ee} maximization problem with multiple users is particularly difficult~\cite{zappone2015energy}, the proposed solution is expected to have good generalizability to other power control problems.

\begin{change}

In the literature,
multiple globally optimal but complicated methods are proposed to optimize the \gls{ee}.
For example,
fractional programming~\cite{isheden2012framework,ng2012energy2,he2013coordinated,zappone2015energy,zappone2017globally},
successive incumbent transcending~\cite{matthiesen2019efficient},
branch-and-bound~\cite{matthiesen2020globally},
dual problem~\cite{ng2012energy},
iterative water filling and exhaustive search~\cite{xu2013energy},
block coordinate descent~\cite{alonzo2019energy}
are applied to optimize the \gls{ee}.
Due to the high complexity (e.g., the complexity of fractional programming grows exponentially with number of links~\cite{zappone2017globally}),
the above-mentioned algorithms are difficult to be applied in real time.
On the other hand, less complex and suboptimal methods are proposed, including
\gls{sca}~\cite{cai2020joint},
sequential fractional programming~\cite{zappone2017globally},
dual Dinkelbach~type algorithm~\cite{du2014distributed},
binary quantum-behaved particle swarm optimization (BQPSO)~\cite{xu2014energy}.
However, the low complexity is achieved with simplification and approximation and the performance is suboptimal~\cite{zappone2017globally},
or with strong assumptions (such as effective interference cancellation, i.e., noise-limited networks~\cite{zappone2015energy})
and the application is limited.
An attempt to realize a good balance between complexity and performance is to apply supervised learning,
i.e., to let a \gls{nn} \enquote{memorize} the optimal power allocation obtained with the above-mentioned complex and optimal analytical methods for massive data.
According to the universal approximation theorem~\cite{hornik1989multilayer,elbrachter2021deep}, deep neural networks can well approximate any continuous function in high dimensional space.
This property allows for higher potential of performance than analytical methods with suboptimal approximations.
These efforts include \cite{matthiesen2020globally,zappone2018online,zhang2019energy,besser2020deep}.
Although the application of the \gls{nn} facilitates the real-time application
because applying the trained \gls{nn} to new inputs is simple and fast,
applying the complex analytical methods to massive data for label generation
is necessary
and the overall computation effort is still high.
Besides, the usability and scalability of the trained \glspl{nn} are limited by the analytical methods
and it is less exciting to let the \gls{nn} only \enquote{memorize} the correct answer.
It is also proposed to let the \gls{nn} try to find the optimal power allocation with unsupervised learning~\cite{chang2018joint,huang2022unsupervised} and reinforcement learning~\cite{lee2020deep,mastronarde2011fast,ram2022energy,omoniwa2022optimizing},
which does not require \enquote{correct} answers as labels but lets the algorithm find the optimal transmit power by itself.
Therefore, the requirement on training data and computational power is considerably lower than supervised learning.
However, these works adopt model-free learning (as explained in detail below)
or have not taken the two most important properties of the problem, permutation-equivariance and non-convexity (as described in Section~\ref{sec:problem}) fully into consideration and the performance is not optimal.

Considering the limitations of the above-mentioned works,
we propose an unsupervised, model-based machine learning approach to solve the \gls{ee} maximization problem
(unlike the model-free reinforcement learning approaches~\cite{lee2020deep,mastronarde2011fast,ram2022energy,omoniwa2022optimizing}).
In model-based machine learning,
the optimizer knows the objective function and computes its gradient to optimize the \gls{nn} 
(on the contrary, the model-free machine learning optimizes the \gls{nn} based on feedback from unknown mechanism 
in a trial-and-error manner~\cite{de2005tutorial}).
Since the model-based learning has more information about the objective than the model-free learning,
it usually has a higher sample efficiency~\cite{nagabandi2018neural}.
Compared to these existing machine learning methods mentioned above,
our proposed approach has the following two innovations:
\begin{itemize}
\item
We propose an innovative objective function, which replaces the original \gls{ee} expression in the gradient ascent method to overcome the non-convexity problem (Section~\ref{sec:rep}).
The non-convexity is a long-existing problem in machine learning with possible counter-measures such as alternating optimization, \gls{em} algorithm, dual problem and stochastic optimization~\cite{jain2017non}.
Our proposed solution is based on the stochastic optimization and the reparameterization trick, which approximates the expectation of the objective with the empirical mean of random variables.
The reparameterization trick makes the random variable differentiable and therefore enables \gls{sga}.
It is shown that this approach has a smaller variance compared to other methods~\cite{xu2019variance}.
In recent years, it has been successfully applied to Bayesian optimization~\cite{wilson2017reparameterization}, variational auto-encoder~\cite{caterini2018hamiltonian} and reinforcement learning~\cite{haarnoja2018soft},
but not yet in non-convex unsupervised learning, in particular, not for the interference network optimization problem.
\item
We design a permutation-equivariant \gls{nn} architecture \InterferenceNet, i.e., if the input pairs of user and \gls{bs} are permuted,
the output powers of the \gls{nn} are permuted automatically in the same way~\cite{gordon2019permutation,ravanbakhsh2017equivariance} (Section~\ref{sec:nic}).
This increases the generality of the \gls{nn} significantly, since the permutation-equivariance is an inherent property of the power control problems in cellular networks.
However, classical \gls{nn} architectures, such as \glspl{nn} with fully connected layers, do not have this property.
In the literature, 
a dedicated \gls{nn} architecture \emph{PowerNet} is proposed for power control~\cite{zhang2019energy} (however, it is not permutation-equivariant),
a permutation-equivariant \gls{nn} architecture has been applied to signal detection~\cite{pratik2020re},
and
the permutation-equivariant \gls{gnn} has been applied to capacity maximization (different from our objective)~\cite{eisen2020optimal,wang2022learning,chowdhury2021unfolding}.
However, these works are not particularly optimized for the non-convexity of the considered problem and the performance is not compared with the global optimum.
\end{itemize}
\end{change}

\emph{Notation:}
A normal small letter (e.g., $s$) denotes a scalar,
a bold small letter (e.g., $\mathbf{v}$) denotes a vector,
a bold capitalized letter (e.g., $\mathbf{M}$) denotes a matrix or a tensor of dimension higher than 2,
$\mathbf{M}^T$ denotes the transpose of matrix~$\mathbf{M}$,
$|\mathcal{S}|$ denotes the cardinality of set~$\mathcal{S}$,
$\mathrm{dim}(\mathbf{v})$ denotes the number of dimensions of vector~$\mathbf{v}$,
$\text{diag}(\mathbf{M})$ denotes the diagonal elements of matrix~$\mathbf{M}$,
$\mathbf{I}$ denotes the identity matrix,
$\mathbf{0}$ denotes the matrix of all zeros and
$\mathbf{1}$ denotes the matrix of all ones,
the extraction matrix~$\mathbf{E}$ is defined as $\mathbf{E} = \mathbf{1} - \mathbf{I}$,
$\mathcal{U}(\mathbf{a}, \mathbf{b})$ denotes a uniform distribution between $\mathbf{a}$ and $\mathbf{b}$ where the $i$-th component is uniformly distributed between $a_i$ and $b_i$ and all components in the distribution are independent from each other,
$\odot$ denotes the element-wise product of two vectors,
$\prod\limits^\circ \mathbf{v}$ denotes the product of elements of $\mathbf{v}$,
$\sum\limits^\circ \mathbf{v}$ denotes the sum of elements of $\mathbf{v}$,
$\mathbf{T}[i, j, k]$ denotes element in position $(i, j, k)$ in tensor $\mathbf{T}$,
$\mathbf{T}[\cdot, j, k]$ denotes the vector of the elements with position $(j, k)$ in the second and third dimensions in tensor $\mathbf{T}$,
$\mathbf{T}[i, \cdot, \cdot]$ denotes the matrix of the elements with position $i$ in the first dimension in tensor $\mathbf{T}$,
$\mathbf{A} \bullet \mathbf{B}$ denotes multiplication between matrix~$\mathbf{A}$ and tensor~$\mathbf{B}$,
where $\mathbf{A}$ is multiplied with all vectors $\mathbf{B}[\cdot, i, j]$ for all $i$ and $j$,
$\mathbf{A} \circ \mathbf{B}$ denotes multiplication between matrix~$\mathbf{A}$ and tensor~$\mathbf{B}$,
where $\mathbf{A}$ is multiplied with all matrices $\mathbf{B}[i, \cdot, \cdot]$ for all $i$,
or multiplication between tensor $\mathbf{A}$ and matrix $\mathbf{B}$,
where all matrices $\mathbf{A}[i, \cdot, \cdot]$ (for all $i$) is multiplied with $\mathbf{B}$.

\section{Problem Statement}
\label{sec:problem}

\subsection{Problem Formulation}
\begin{figure}
    \centering
    \resizebox{.2\textwidth}{!}{
    \begin{tikzpicture}
\node (bs1) at (0,0) {\includegraphics[height=30pt]{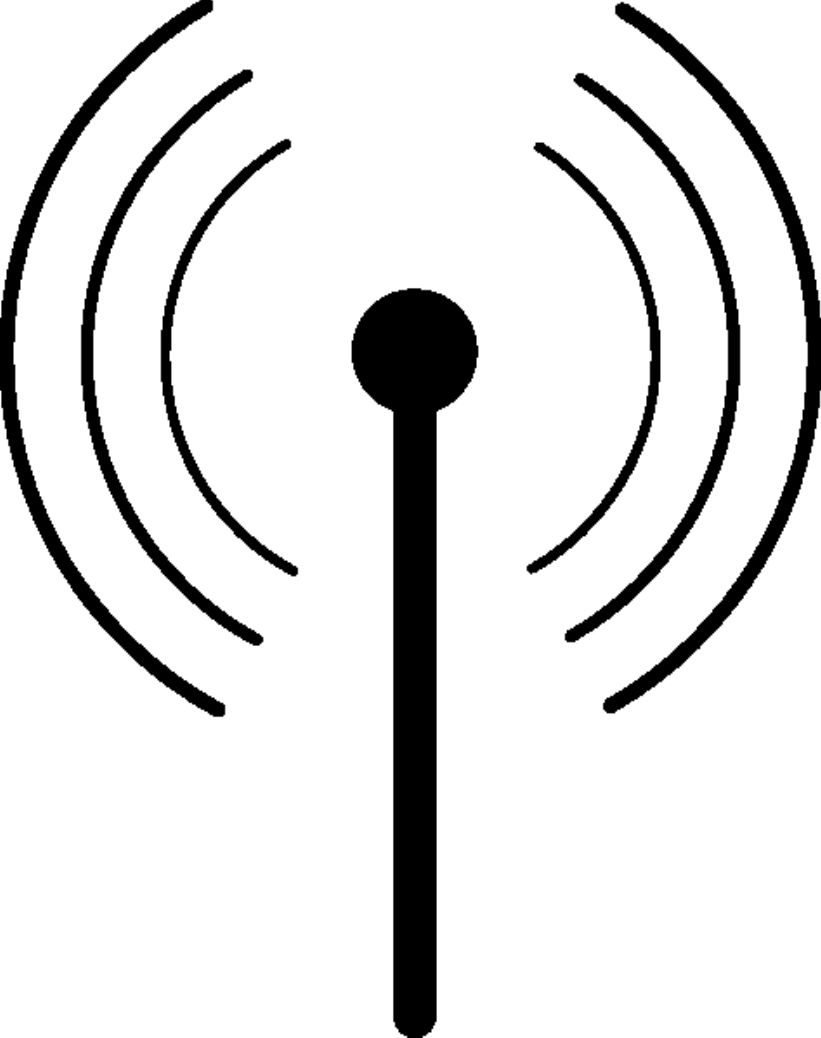}};
\node (bs2) at (4,0) {\includegraphics[height=30pt]{figs/bs.pdf}};
\filldraw[black] (4.4,3.7) circle (0pt) node[anchor=west]{BS~$i$};
\node (bs3) at (4,4) {\includegraphics[height=30pt]{figs/bs.pdf}};
\node (bs4) at (0,4) {\includegraphics[height=30pt]{figs/bs.pdf}};

\node (ue1) at (1, -0.4) {\includegraphics[height=20pt]{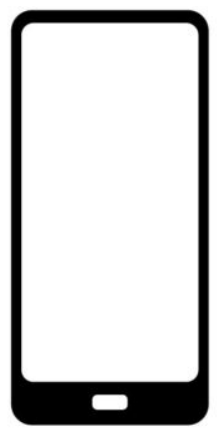}};
\node (ue2) at (-1, 0.4) {\includegraphics[height=20pt]{figs/ue.pdf}};

\node (ue3) at (5, 0.6) {\includegraphics[height=20pt]{figs/ue.pdf}};

\node (ue4) at (0.3, 3.3) {\includegraphics[height=20pt]{figs/ue.pdf}};
\node (ue5) at (0.5, 4.9) {\includegraphics[height=20pt]{figs/ue.pdf}};

\node (ue6) at (5.5, 4.7) {\includegraphics[height=20pt]{figs/ue.pdf}};
\filldraw[fill=white] (4.5,6.0) circle (0pt) node[anchor=south]{User~$i$};
\node (ue7) at (4.5, 5.7) {\includegraphics[height=20pt]{figs/ue.pdf}};

\draw[->, dotted] (1, -0.4) -- (4, 4);
\draw[->, dotted] (-1, 0.4) -- (4, 4);
\draw[->, dotted] (5, 0.6) -- (4, 4);
\draw[->, dotted] (0.3, 3.3) -- (4, 4);
\draw[->, dotted] (0.5, 4.9) -- (4, 4);
\draw[->, dotted] (5.5, 4.7) -- (4, 4);
\draw[->] (4.3, 5.3) -- (4, 4);
\end{tikzpicture}}
    \caption{Illustration of the considered scenario. The channel of the signal of interest for the user-\gls{bs} pair~$i$ is drawn as the solid line and the interference channels are drawn as the dotted lines.}
    \label{fig:scenario}
\end{figure}

We consider an uplink transmit power control problem in a multi-cell wireless network,
as shown in Figure~\ref{fig:scenario},
where $I$ single-antenna users are served by $M$ \glspl{bs} equipped with $n_R$ antennas.
We assume that every user is assigned to one \gls{bs} and denote the assigned \gls{bs} of user~$i$ as \gls{bs}~$i$.
We further assume that the \glspl{bs} use the matched filter to process the received signal.
User~$i$ chooses a transmit power~$p_i \in [0, p_{\max}]$,
where $p_{\max}$ is the maximum transmit power\footnote{Information available at the user is usually limited in practice,
which might be insufficient to obtain the global optimum.
The \gls{bs} can choose a $p_i$ based on the complete information available at the \gls{bs} and inform the user.}.
\begin{change}
Our objective is to maximize the sum-\gls{ee},
which is defined as the sum of ratios between data rate and power consumption of all user-\gls{bs} pairs.
This objective is inherited from \cite{matthiesen2020globally} and is an indicator of the overall \gls{ee} of the considered multi-cell communication system.
The summation suggests that we do not favor any link in particular.
In fact,
if all transmitters are energy-limited devices,
their individual \gls{ee} matters and
the sum \gls{ee} is a more suitable metric compared to the global \gls{ee} (sum of data rates divided by sum of energy consumption).
\end{change}
The problem is formulated as
\begin{equation}
\begin{aligned}
    \max_{\mathbf{p}} \quad & J = \sum_{i=1}^I \frac{\log\left(1 + \frac{g_{ii} p_i}{1 + \sum_{j \neq i} g_{ji} p_j}\right)}{\mu p_i + P_c}\\
    \text{s.t.} \quad & 0 \leq p_i \leq p_{\max} \quad \text{ for } i = 1, 2, \dots, I,
\end{aligned}
\label{eq:problem}
\end{equation}
where $\mathbf{p} = ( p_1, p_2, \dots, p_I )$ is the vector of transmit powers of all user-\gls{bs} pairs,
$g_{ji}$ is the equivalent channel gain from user~$j$ to \gls{bs}~$i$ assuming the matched filter receiver.
Therefore, $g_{ii}$ is the channel gain of the useful signal and $g_{ji}$, $j \neq i$, is the channel gain of the interference from user~$i$ to \gls{bs}~$j$,
$\mu$ is the inefficiency of the power amplifier and
$P_c$ is the static power assumption.

\begin{change}
\begin{remark}
The equivalent channel gains $g_{ij}$ is assumed known in this problem,
which is a widely used and practical assumption~\cite{he2013coordinated,zappone2015energy,matthiesen2020globally,besser2020deep}
since we do not assume many antennas (such as massive MIMO and reconfigurable intelligent surface).
The channel estimation can be performed via pilot signals,
whose energy consumption is not considered in this work for two reasons:
1) The pilot signal and its energy consumption is far less than data traffic signal~\cite{pilot}.
2) The energy consumption of the pilot signal is independent from the power control strategy for the traffic signal.
Therefore, the power optimization of channel estimation is decoupled with the power optimization of data traffic.
Besides, there are also models of statistical \gls{csi} where the resulting expressions have the same structure as with perfect \gls{csi}.
This is achieved by pulling the expectation operator in the denominator and numerator of the \gls{sinr} expression separately.
This is one possible extension of this work.
\end{remark}
\end{change}

\subsection{Properties of the Problem}
\subsubsection{Non-convexity}
A non-convex function is a function whose second-order derivative is not non-negative on its entire domain.
If a function is non-convex, it can have multiple local optima.
From \eqref{eq:problem}, we notice that the objective~$J$ is a non-convex function in power $\mathbf{p}$.
The conventional convex optimization methods might converge to a poor local optimum, if we use them to optimize a non-convex objective.
In the literature, \cite{matthiesen2020globally} proposes to solve the problem with a branch-and-bound algorithm,
which guarantees to converge to the global optimum at the cost of high computational effort and poor scalability.
Alternatively, in standard \gls{nn} optimization, we use gradient ascent/descent to optimize the \gls{nn},
which cannot not be prevented from converging to a local optimum.

\subsubsection{Permutation-Equivariance}
\label{sec:equivariance}
Denote the equivalent channel matrix as $\mathbf{G}$, where the element in row~$i$ and column~$j$ is $g_{ij}$ as defined in \eqref{eq:problem}.
We expect that the power allocation $\mathbf{p}$ is permutation-equivariant to $\mathbf{G}$,
i.e.,
if the user-\gls{bs} pairs are permuted,
the elements in $\mathbf{p}$ should be permuted in the same way.
This is because the order of the user-\gls{bs} pairs is inherently arbitrary and should have no impact on the optimal transmit power.
Mathematically, we use the permutation matrix $\mathbf{P}$ to describe a permutation,
whose elements are either 0 or 1 and
the sum of every row and every column is 1.
For example, the matrix
\begin{equation*}
    \mathbf{P} =
    \begin{pmatrix}
         1 & 0 & 0 \\
         0 & 0 & 1 \\
         0 & 1 & 0
    \end{pmatrix}
\end{equation*}
is a valid permutation matrix.
The product~$\mathbf{P} \mathbf{G}$ permutes the rows of matrix $\mathbf{G}$ whereas $\mathbf{G} \mathbf{P}^T$ permutes the columns of $\mathbf{G}$.
With the example $\mathbf{P}$ defined above,
it permutes the second and third rows (columns) of $\mathbf{G}$ while the first row (column) is unchanged.
If we apply both row and column permutation, the resulting matrix
\begin{equation}
\mathbf{G}' = \mathbf{P} \mathbf{G} \mathbf{P}^T
\label{eq:permutation_matrix}
\end{equation}
is permuted in both rows and columns in the same way simultaneously compared to $\mathbf{G}$.
Consider a mapping~$\Phi: \reals^{I\times I}\to\reals_{+}^{I}$ with $\Phi(\mat{G})=\vec{p}$, 
$\Phi$ is called permutation-equivariant if
\begin{equation}
    \mathbf{p} \mathbf{P}^T = \Phi(\mathbf{P} \mathbf{G} \mathbf{P}^T)
\end{equation}
holds for any permutation matrix $\mathbf{P}$.
Obviously, a conventional \gls{nn} with fully connected layers is not permutation-equivariant.

\begin{change}
\begin{remark}
Compared to problem~\ref{eq:problem},
\gls{sic} is not permutation-equivariant because the decoding order has a direct impact on the performance.
\end{remark}
\end{change}


\section{Application of Reparametrization Trick to Non-Convex Optimization Objectives}
\label{sec:rep}

In this section, we propose an innovative objective function with stochastic actions and reparameterization trick that can allow an optimization to converge to the global optimum even in non-convex problems.

\subsection{Defining the Objective}
\label{sec:objective}
Given a non-convex objective function $J(\mathbf{p})$,
we can maximize $J$ with gradient ascent, where $\mathbf{p}$ is updated as $\mathbf{p} \leftarrow \mathbf{p} + r\nabla_\mathbf{p} J(\mathbf{p})$,
with learning rate $r$ and the gradient of $J(\mathbf{p})$ with respect to $\mathbf{p}$ $\nabla_\mathbf{p} J(\mathbf{p})$.
Since $J$ is non-convex,
there may exist multiple local optima.

\begin{figure}
    \centering
    \resizebox{.3\textwidth}{!}{
    \begin{tikzpicture}[scale=.5]
    \coordinate (y) at (0,3);
    \coordinate (x) at (10,0);
    \draw[<->] (y) node[above] {$J$} -- (0,0) --  (x) node[right]
    {$p$};

    \draw plot [smooth] coordinates { (0, 0.1) (2, 1.4) (5, 0.2) (7, 3) (8.5, 0.4) (10, 0.2)};
    \draw[fill] (1.4, 1.2) circle [radius=0.1];
    \draw[fill] (2, 1.4) circle [radius=0.1];
    \draw[fill] (7, 3) circle [radius=0.1];
    \draw [dashed] (1.4, 0) -- (1.4, 1.2);
    \node at (1.4, -0.4) {$p^{(0)}$};
    \node at (2, 1.8) {$O_L$};
    \node at (7, 3.4) {$O_G$};
    \end{tikzpicture}}
    \caption{Illustration of a non-convex objective function~$J$. When $p$ is initialized as $p^{(0)}$ near the local optimum~$O_L$, the classical gradient ascent optimization converges to the local optimum~$O_L$, which is not the global optimum~$O_G$.}
    \label{fig:non-convex}
\end{figure}
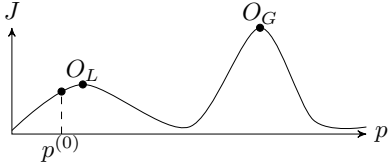

In Figure~\ref{fig:non-convex}, it is illustrated that
if $p$ is initialized as $p^{(0)}$ near a poor local optimum $O_L$,
the optimization with gradient ascent would converge to $O_L$ rather than the global optimum $O_G$.
Note that we reduced the vector~$\vec{p}$ to a one-dimensional scalar~$p$ for illustration simplicity.

In order to solve this problem,
we do not use a deterministic $\mathbf{p}$.
Instead, we assume that $\mathbf{p}$ is a random variable
with $\mathbf{p} \sim \mathcal{U}(\mathbf{a}, \mathbf{b})$, where $\mathbf{a}$ and $\mathbf{b}$ are lower and upper bound of the support of the distribution, respectively.
The expectation of the objective~$J$ is
\begin{equation}
    K = \mathbb{E}_{\mathbf{p} \sim \mathcal{U}(\mathbf{a}, \mathbf{b})}(J\left(\mathbf{p})\right).
    \label{eq:objective1}
\end{equation}
We can use the mean of multiple \gls{iid} random variables to approximate the expectation in \eqref{eq:objective1}.
However, we cannot compute the gradient of a random variable and gradient ascent cannot be applied to optimize $K$ as a result.
Therefore, we use the reparameterization trick.
Instead of sampling $\mathbf{p}$ from the distribution $\mathcal{U}(\mathbf{a}, \mathbf{b})$ directly,
we sample $\mathbf{p}$ from the standard uniform distribution $\mathcal{U}(\mathbf{0}, \mathbf{1})$
and rewrite \eqref{eq:objective1} as
\begin{equation}
    K = \mathbb{E}_{\mathbf{p} \sim \mathcal{U}(\mathbf{0}, \mathbf{1})}(J\left(\mathbf{a} + (\mathbf{b} - \mathbf{a}) \odot \mathbf{p})\right),
    \label{eq:objective1.1}
\end{equation}
where we use the obvious fact $\mathbf{a} + (\mathbf{b} - \mathbf{a}) \odot \mathbf{p} \sim \mathcal{U}(\mathbf{a}, \mathbf{b})$ if $\mathbf{p} \sim \mathcal{U}(\mathbf{0}, \mathbf{1})$.
In this way, we can sample $\mathbf{p}$ from a fixed distribution to approximate the expectation,
compute $\nabla_{\mathbf{a}}K$ and $\nabla_{\mathbf{b}}K$ and
perform a gradient ascent step to improve $K$.

It would be beneficial for the optimization to converge to the global optimum if the global optimum lies within $[\mathbf{a}, \mathbf{b}]$.
Therefore, it is desirable that $\mathbf{a}$ and $\mathbf{b}$ are initialized outside the feasible region.
It is then necessary to introduce penalty terms 
\begin{equation}
    P(\mathbf{a}) = 
    -\min\left(\sum\limits^\circ\left(\mathbf{a} - \mathbf{p}_{\min}\right), 0\right)
    \label{eq:penalty1}
\end{equation}
and
\begin{equation}
    Q(\mathbf{b}) = 
    \max\left(\sum\limits^\circ\left(\mathbf{b} - \mathbf{p}_{\max}\right), 0\right),
    \label{eq:penalty2}
\end{equation}
where $\mathbf{p}_{\min}$ and $\mathbf{p}_{\max}$ define the feasible region of the power allocation.
The penalty terms \eqref{eq:penalty1} and \eqref{eq:penalty2} are 0 if $\mathbf{a}$ or $\mathbf{b}$ are inside the feasible region,
otherwise they are proportional to the distances from the feasible region.

\begin{figure}
    \centering
    \subfigure[In the beginning of optimization]{%
    \label{fig:nonconvex_objective_begin}
    \centering
    \resizebox{.3\textwidth}{!}{
    \begin{tikzpicture}[scale=.5]
    \coordinate (y) at (0,3);
    \coordinate (x) at (10,0);
    \draw[<->] (y) node[above] {$J$} -- (0,0) --  (x) node[right]
    {$p$};

    \draw plot [smooth] coordinates { (0, 0.1) (2, 1.4) (5, 0.2) (7, 3) (8.5, 0.4) (10, 0.2)};
    \draw [dashed] (0.5, 0) -- (0.5, 0.5);
    \node at (0.5, -0.4) {$a$};
    \draw[fill] (0.5,0.5) circle [radius=0.1];
    \draw [dashed] (8.35, 0) -- (8.35, 0.45);
    \node at (8.35, -0.4) {$b$};
    \draw [dashed] (0, 1.1) -- (10, 1.1);
    \node at (10, 1.6) {$\mathbb{E}_{\mathbf{p} \sim \mathcal{U}(a, b)}(J)$};
    \draw[fill] (8.35,0.6) circle [radius=0.1];
    \end{tikzpicture}
    }
    }
    \subfigure[No entropy regularization required]{%
    \label{fig:nonconvex_objective-above-local}
    \centering
    \resizebox{.3\textwidth}{!}{
    \begin{tikzpicture}[scale=.5]
    \coordinate (y) at (0,3);
    \coordinate (x) at (10,0);
    \draw[<->] (y) node[above] {$J$} -- (0,0) --  (x) node[right]
    {$p$};

    \draw plot [smooth] coordinates { (0, 0.1) (2, 1.4) (5, 0.2) (7, 3) (8.5, 0.4) (10, 0.2)};
    \draw [dashed] (1, 0) -- (1, 0.9);
    \node at (1, -0.4) {$a$};
    \draw[fill] (1,0.9) circle [radius=0.1];
    \draw [dashed] (8.05, 0) -- (8.05, 1.2);
    \node at (8.05, -0.4) {$b$};
    \draw [dashed] (0, 1.2) -- (10, 1.2);
    \draw[fill] (8.05,1.2) circle [radius=0.1];
    \node at (10, 1.6) {$\mathbb{E}_{\mathbf{p} \sim \mathcal{U}(a, b)}(J)$};
    \end{tikzpicture}
    }
    }
    \subfigure[Entropy regularization required]{%
    \label{fig:nonconvex_objective-below-local}
    \centering
    \resizebox{.3\textwidth}{!}{
    \begin{tikzpicture}[scale=.5]
    \coordinate (y) at (0,3);
    \coordinate (x) at (10,0);
    \draw[<->] (y) node[above] {$J$} -- (0,0) --  (x) node[right]
    {$p$};

    \draw plot [smooth] coordinates { (0, 0.1) (2, 1.4) (5, 0.2) (7, 3) (8.5, 0.4) (10, 0.2)};
    \draw [dashed] (1.4, 0) -- (1.4, 1.2);
    \node at (1.4, -0.4) {$a$};
    \draw[fill] (1.4, 1.2) circle [radius=0.1];
    \draw [dashed] (8.05, 0) -- (8.05, 1.2);
    \node at (8.05, -0.4) {$b$};
    \draw [dashed] (0, 1.2) -- (10, 1.2);
    \draw[fill] (8.05, 1.2) circle [radius=0.1];
    \node at (10, 1.6) {$\mathbb{E}_{\mathbf{p} \sim \mathcal{U}(a, b)}(J)$};
    \end{tikzpicture}
    }
    }
    \caption{A non-convex objective function and its mean in the interval $[a, b]$.
    (a): Both $J(a)$ and $J(b)$ are less than $\mathbb{E}_{\mathbf{p} \sim \mathcal{U}(a, b)}(J)$,
    performing a gradient ascent step of $\mathbb{E}_{\mathbf{p} \sim \mathcal{U}(a, b)}(J)$ increases $a$ and decreases $b$.
    (b): After a few iterations,
    $J(a) < \mathbb{E}_{\mathbf{p} \sim \mathcal{U}(a, b)}(J)$ and $J(b) = \mathbb{E}_{\mathbf{p} \sim \mathcal{U}(a, b)}(J)$.
    A gradient ascent step increases $a$ and does not change $b$.
    (c): Now we have $J(a) = J(b) = \mathbb{E}_{\mathbf{p} \sim \mathcal{U}(a, b)}(J)$.
    Performing gradient ascent changes neither $a$ nor $b$.
    We need the entropy regularization term to help $a$ to \enquote{cross over} the local optimum.}
    \label{fig:nonconvex_objective}
\end{figure}
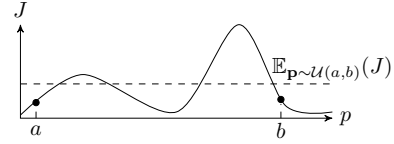
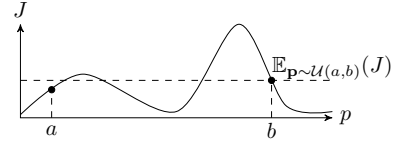
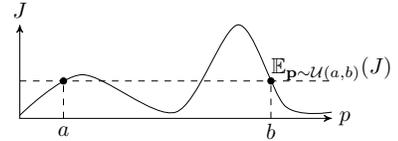

In one iteration,
if $J(a) < \mathbb{E}_{\mathbf{p} \sim \mathcal{U}(a, b)}(J)$ and $J(b) < \mathbb{E}_{\mathbf{p} \sim \mathcal{U}(a, b)}(J)$,
as shown in Figure~\ref{fig:nonconvex_objective_begin},
a gradient ascent step will increase $a$ and decrease $b$.
If $J(a) < \mathbb{E}_{\mathbf{p} \sim \mathcal{U}(a, b)}(J)$ and $J(b) = \mathbb{E}_{\mathbf{p} \sim \mathcal{U}(a, b)}(J)$,
as shown in Figure~\ref{fig:nonconvex_objective-above-local},
$\mathbf{a}$ increases and $\mathbf{b}$ is unchanged in a gradient ascent step.
If $J(a) = J(b) = \mathbb{E}_{\mathbf{p} \sim \mathcal{U}(a, b)}(J)$,
as shown in Figure~\ref{fig:nonconvex_objective-below-local},
neither increasing $\mathbf{a}$ nor decreasing $\mathbf{b}$ improves $\mathbb{E}_{\mathbf{p} \sim \mathcal{U}(\mathbf{a}, \mathbf{b})}(J)$.
In this case, $\mathbf{a}$ and $\mathbf{b}$ cannot converge if we use \eqref{eq:objective1.1} as the objective function.
Therefore, we introduce an entropy regularization term in the objective function as an additional incentive to reduce the entropy.
For multivariate uniform distribution with independent components, its entropy is given as
\begin{equation}
    H(\mathbf{a}, \mathbf{b}) = \ln \left(\prod\limits^\circ(\mathbf{b} - \mathbf{a})\right).
    \label{eq:entropy}
\end{equation}
The total objective function~$L$, including the penalties and the entropy regularization, is
\begin{equation}
\begin{aligned}
    L = & \mathbb{E}_{\mathbf{p} \sim \mathcal{U}(\mathbf{0}, \mathbf{1})} J(\mathbf{a} + (\mathbf{b}
    - \mathbf{a})\odot\mathbf{p}) \\
    & - \epsilon P(\mathbf{a})
    - \epsilon Q(\mathbf{b})
     - \kappa H(\mathbf{a}, \mathbf{b}),
\end{aligned}
\label{eq:objective2}
\end{equation}
where $\epsilon$ is the Lagrange multiplier for support region constraint
and $\kappa$ is the coefficient for the entropy regularization term,
whose value needs to be tuned during the optimization process,
which is discussed in Section~\ref{sec:entropy}. 
If we compute the gradients of \eqref{eq:objective2} with respect to $\mathbf{a}$ and $\mathbf{b}$, i.e., $\nabla_{\mathbf{a}} L$ and $\nabla_{\mathbf{b}} L$,
perform gradient ascents $\mathbf{a} \leftarrow \mathbf{a} + r \cdot \nabla_{\mathbf{a}} L$ and $\mathbf{b} \leftarrow \mathbf{b} + r \cdot \nabla_{\mathbf{b}} L$,
where $r$ is the learning rate, 
and repeat the process iteratively
until the entropy of $\mathcal{U}(\mathbf{a}, \mathbf{b})$ is smaller than a predefined threshold $H_0$,
$\mathbf{a}$ and $\mathbf{b}$ have a high possibility to converge to the global optimum.

\begin{figure}
    \centering
    \resizebox{.3\textwidth}{!}{
    \begin{tikzpicture}[scale=.5]
    \coordinate (y) at (0,3);
    \coordinate (x) at (10,0);
    \draw[<->] (y) node[above] {$J$} -- (0,0) --  (x) node[right]
    {$p$};

    \draw plot [smooth] coordinates { (0, 0.1) (0.9, 0.68) (1, 1.7) (1.1, 0.72) (2, 1) (5, 0.2) (7, 3) (8.5, 0.4) (10, 0.2)};
    \draw [dashed] (1, 0) -- (1, 1.5);
    \node at (1, -0.4) {$a$};
    \draw[fill] (1, 1.7) circle [radius=0.1];
    \draw [dashed] (7.5, 0) -- (7.5, 2.45);
    \draw[fill] (7.5, 2.45) circle [radius=0.1];
    \node at (7.5, -0.4) {$b$};
    \end{tikzpicture}}
    \caption{An objective function with a flat global optimum and a sharp local optimum.}
    \label{fig:sharp_local}
\end{figure}
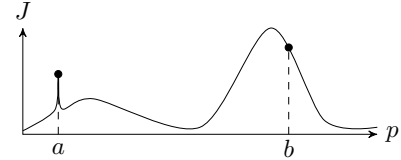

\begin{remark}
\label{remark:kappa}
Note that the gradient of the integral of $J$ is $J$ itself.
Therefore, whether $\mathcal{U}(\mathbf{a}, \mathbf{b})$ reduces in a gradient ascent step
depends on the values $J(\mathbf{G}, \mathbf{a})$ and $J(\mathbf{G}, \mathbf{b})$ themselves,
not on their gradients.
A flat global optimum is therefore more advantageous than a sharp local optimum.
In the case of Figure~\ref{fig:sharp_local},
$\nabla_a \mathbb{E}_{\mathbf{p} \sim \mathcal{U}(a, b)}(J) < \nabla_b \mathbb{E}_{\mathbf{p} \sim \mathcal{U}(a, b)}(J)$ because $J(a) < J(b)$.
With a properly chosen $\kappa$,
$\mathbf{a}$ can \enquote{jump} over the sharp and poor local optimum and
$\mathbf{b}$ stays in front of the global optimum.
\end{remark}

\subsection{Tuning the Lagrange Multiplier for Entropy Term}
\label{sec:entropy}
From Section~\ref{sec:objective}, we realize that tuning $\kappa$ in \eqref{eq:objective2} is crucial to the success of the training 
because it has to be large enough such that the training does not get stuck at local optima
and it has to be small enough such that the global optimum stays in the support of $\mathcal{U}(\mathbf{a}, \mathbf{b})$.

In the literature, it is suggested to optimize the factor of the entropy term via the dual problem for a similar original problem~\cite{haarnoja2018soft}.
However, since it is very difficult to obtain the infimum of the Lagrangian of our complicated objective function (unlike the canonical definition of the dual problem),
there is no guarantee of convexity of the dual problem,
which would be a fatal disadvantage in the considered problem.
We propose a simple heuristic approach to tune $\kappa$ in this section. 
Let $\kappa_i$ be the $\kappa$ in iteration~$i$ of the gradient ascent optimization.
$\kappa_i = 0$ for $i \leq h$,
where $h$ is a hyperparameter for constant $\kappa$.
For $i > h$,
$\kappa$ is computed as
\begin{equation}
    \kappa_{i + 1} =
    \begin{cases}
    \kappa_i + \Delta\kappa & \text{if } \sum_{j = 1}^{h} H_{i - j} / h \leq H_i \\
    \max(0, \kappa_t - \Delta\kappa / 2) & \text{otherwise,}\\
    \end{cases}
    \label{eq:updating_kappa}
\end{equation}
where $\Delta\kappa$ is a constant small learning rate.
The intuition behind \eqref{eq:updating_kappa} is that we carefully increase $\kappa$ if the entropy does not reduce (case~1)
and decrease $\kappa$ otherwise such that $\kappa$ is not too big to make $\mathbf{a}$ or $\mathbf{b}$ cross the global optimum (case~2).

\begin{figure}[htbp]
    \centering
    \input{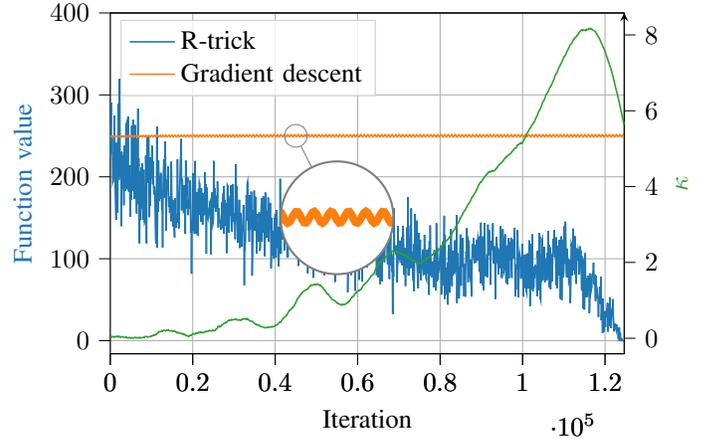}
    \caption{Optimization of the Rastrigin function with the proposed method and gradient descent.}
    \label{fig:ackley}
\end{figure}

\begin{example*}
In order to demonstrate the ability of the proposed objective function,
we apply the widely-used Rastrigin function for performance testing~\cite{rudolph1990globale},
which is the overlapping of a parabolic function and cosine functions:
\begin{equation}
     f(\mathbf{x}) = An + \sum_{i=1}^n (x_i^2 - A\cos(2\pi x_i)),
\label{eq:rastrigin}
\end{equation}
where $A=10$ and $n$ is the number of dimensions.
$f(\mathbf{x})$ is non-convex with many local minima and its global minimum is $f(0, 0, \dots, 0) = 0$.
We assume $n=10$ and use the proposed method \eqref{eq:objective2} and conventional gradient descent to minimize \eqref{eq:rastrigin}.
$\kappa$ is initialized as 0 
and penalty terms \eqref{eq:penalty1} and \eqref{eq:penalty2} are omitted because the function is defined for all real numbers.
The results are shown in Figure~\ref{fig:ackley},
while the conventional gradient descent gets stuck in a poor local minimum,
the proposed method converges to the global optimum successfully.
This result demonstrates the ability of the proposed method to find the global optimum of a simple non-convex objective function.
\end{example*}

In principle, we can apply \eqref{eq:objective2} to the power control problem \eqref{eq:problem} to obtain the globally optimal transmit powers $\mathbf{p}$.
However, its high computation effort makes it difficult to be applied in real time
(this fact can be recognized by looking at the number of required iterations in Figure~\ref{fig:ackley}).
Therefore, we do not optimize $\mathbf{a}$ and $\mathbf{b}$ for each channel realization directly.
Instead, we propose to compute the parameters with an \gls{nn} and train the \gls{nn} with massive data.
A new \gls{nn} architecture \emph{\InterferenceNet} is introduced to encode the domain knowledge of interference channels in the \gls{nn} design
and to realize the permutation-equivariance,
as will be discussed in Section~\ref{sec:nic}.


\section{\InterferenceNet: a Permutation-Equivariant Neural Network Architecture for Optimization in Interference Channels}
\label{sec:nic}

In this section, we first introduce the unsupervised learning framework,
then analyze the interference channel model,
propose the \InterferenceNet architecture,
prove its permutation-equivariance.
Finally, we prove its equivalence to \gls{gnn},
which allows for distributed implementation.

\begin{change}

\subsection{Framework of Unsupervised Machine Learning and Comparison to Analytical Methods}
\label{sec:story}

We define $\mathbf{a} = \alpha_\delta(\mathbf{G})$ where $\alpha_\delta$ is an \gls{nn} parameterized by $\delta$ that maps from channel $\mathbf{G}$ to lower bound of the transmit power interval $\mathbf{a}$.
We also define $\mathbf{\ell} = \mathbf{b} - \mathbf{a} = \beta_o(\mathbf{G})$ where $\beta_o$ is an \gls{nn} parameterized by $o$ (omicron)
that maps from channel $\mathbf{G}$ to length of the transmit power interval $\mathbf{\ell}$.
%
Since the objective function~$L$ defined in \eqref{eq:objective2} is fully determined by $\mathbf{G}$, $\mathbf{a}$ and $\mathbf{b}$,
we can write the objective as
\begin{equation}
    L(\mathbf{G}, \mathbf{a}, \mathbf{b}) = L(\mathbf{G}, \alpha_\delta(\mathbf{G}), \alpha_\delta(\mathbf{G}) + \beta_o(\mathbf{G});\delta, o).
    \label{eq:unsupervised1}
\end{equation}
Note that the right hand side of \eqref{eq:unsupervised1} emphasizes that $L$ depends on the parameter $\delta$ and $o$ given $\mathbf{G}$.

We collect massive channel data in a training data set $\mathcal{D}$
and formulate the unsupervised machine learning problem as
\begin{equation}
    \max_{\delta, o} \sum_{\mathbf{G} \in \mathcal{D}} L(\mathbf{G}, \alpha_\delta(\mathbf{G}), \alpha_\delta(\mathbf{G}) + \beta_o(\mathbf{G}); \delta, o).
    \label{eq:ml_objective}
\end{equation}
In this way, we optimize $\alpha_\delta$ and $\beta_o$ which map from any $\mathbf{G} \in \mathcal{D}$ to $\mathbf{a}$ and $\mathbf{b}$, respectively.
The optimization is performed with, e.g., \gls{sga}.
This optimization process is called \emph{training}.
If the data set is general enough,
we would expect that a channel $\mathbf{G}' \notin \mathcal{D}$,
which, however, has similar order of magnitude to channels in $\mathcal{D}$,
can also be mapped to a good transmit power
(on the contrary, we do not expect the trained model works for channel gains which are 10000 times weaker than channel gains in $\mathcal{D}$ on average since there is no extrapolation performance guarantee).
The performance evaluation of $L(\mathbf{G}', \alpha_\delta(\mathbf{G}'), \alpha_\delta(\mathbf{G}') + \beta_o(\mathbf{G}'))$ for $\mathbf{G}' \notin \mathcal{D}$ and trained and fixed $\alpha_\theta$ and $\beta_o$ is called \emph{testing}.
\begin{remark}
Although $\alpha_\delta$ and $\beta_o$ are the same to any $\mathbf{G}$,
different $\mathbf{G}$ result in different $\mathbf{a}$ and $\mathbf{\ell}$.
Therefore, the obtained transmit power is not the average optimal power of all possible channels,
but is optimized for each individual channel.
\end{remark}
\begin{remark}
The training data is not exhaustive.
Therefore, to evaluate the performance of the trained model,
testing with $\mathbf{G}' \notin \mathcal{D}$ is mandatory.
Two keys to ensure the testing performance are 1) the model complexity and training process are reasonable enough to avoid overfitting, 2) the data in $\mathcal{D}$ is general enough such that the trained \gls{nn} is valid for $\mathbf{G}' \notin \mathcal{D}$.
Therefore,
massive data samples are required to ensure the testing performance.
If we meet the two requirements,
the trained \gls{nn} is valid as long as the channel gain distribution does not significantly change (e.g., due to change of frequency band or deployment of femtocell).
\end{remark}
\begin{remark}
Relationship to other algorithms: the supervised learning can be considered as a special case of the proposed training method,
where the objective function is the mean squared error 
$\sum_i(\alpha_\delta(\mathbf{G}_i) - \mathbf{p}_i)^2 / |\mathcal{D}|$
with the $i$th channel data sample $\mathbf{G}_i$ and the $i$th label of transmit power $\mathbf{p}_i$.
The proposed training has also a weak similarity to the deep deterministic policy gradient (DDPG) algorithm in reinforcement learning.
Three main differences are: we make one single decision on transmit power rather than a series of decisions (therefore the Q-value is reduced to reward),
use known \gls{ee} expression instead of learning the Q-value
and apply stochastic decision for the non-convex problem rather than a deterministic one.
\end{remark}

Compare \eqref{eq:objective2} and \eqref{eq:ml_objective},
we notice a fundamental difference between traditional optimization approaches (including the one introduced in Section~\ref{sec:rep}) and machine learning:
instead of optimizing $\mathbf{a}$ and $\mathbf{\ell}$ themselves,
the machine learning approach optimizes the functions (realized by $\alpha_\delta$ and $\beta_o$) that map from channels $\mathbf{G}$ to $\mathbf{a}$ and $\mathbf{\ell}$.
While the function optimization (training) is time-consuming,
the application of the function to a new channel without modifying the function itself (testing and deployment) requires significantly less computational power.
Moreover, the dedicated \gls{nn} architecture \InterferenceNet presented in this section
exploits the fact that the channels are homogeneous and therefore
can utilize the power of parallel computing,
as will be explained in detail later in this section.
This is the reason that the machine learning approach requires less computational power in application than the analytical methods.
\end{change}

\begin{change}
\subsection{From Interference Channel to \InterferenceNet}
\label{sec:from_to}

The underlying information-theoretical model of problem~\eqref{eq:problem} is the interference channel~\cite{carleial1978interference}.
Note that the interference channel is from information sources to their destinations,
including precoding at the transmitter, wireless channel and decoding at the receiver.
Therefore, the number of links in the interference channel only depends on the number of transmitter-receiver pairs and is independent from the number of antennas.
We have the following observations that inspire the design of the \gls{nn} architecture:
\begin{itemize}
    \item There are $I^2$ channels given $I$ user-\gls{bs} pairs,
    which implies that problem \eqref{eq:problem} is high dimensional with an ordinary number of user-\gls{bs} pairs.
    However, the direct and interference channels are homogeneous,
    which suggests that we can apply the same information processing on every channel.
    If we can do this, the complexity of the \gls{nn} is independent from the number of user-\gls{bs} pairs
    and the high dimensionality is no more a constraint.
    Overfitting can be avoided to a large extent.
    Furthermore, we can exploit the power of parallel computing because information processing of different channels are independent from each other.
    \begin{remark}
    On the contrary, many widely-used analytical methods perform alternating optimization and cannot parallelize information processing of antennas (e.g., block coordinate descent (BCD)~\cite{Guo_Liang_Chen_Larsson_2020}) or propagation paths (e.g., space alternating generalized expectation-maximization (SAGE)~\cite{Fessler_Hero_1994}) because they depend on each other.
    \end{remark}
    \item Although there are $I^2$ channels,
    they can all be classified into 4 categories of channels\footnote{Note that channel $c_{ij}$ and channel gain $g_{ij}$ are different: channel $c_{ij}$ refers to the link from transmitter~$j$ to receiver~$i$.
    For example, we say that channel $c_{11}$ and channel $c_{21}$ share the same transmitter~1.
    Channel gain $g_{ij}$ is the ratio between received signal power at receiver~$i$ and transmitted signal power at transmitter~$j$.} for channel $c_{ij}$ from user~$j$ to \gls{bs}~$i$ according to their roles in the \gls{sinr} expression, namely
    \begin{enumerate}
        \item $\mathcal{C}^1(c_{ij}) = \{ c_{ij} \}$, where $\mathcal{C}^d(c_{ij})$ is the set of channels in category~$d$ for channel $c_{ij}$.
        This category contains only the channel $c_{ij}$ itself.
        The corresponding channel gain $g_{ij}$ appears in the numerator of the \gls{sinr} of the corresponding link.
        If $g_{ij}$ is high,
        a high received signal strength is achieved with the same transmit power.
        \item $\mathcal{C}^2(c_{ij}) = \{ c_{kj}, k = 1, \dots, N, k \neq i \}$.
        These are channels that are interfered by the source of channel $c_{ij}$ (user~$j$). 
        Channel gain $g_{ij}$ appears in the denominator of the \gls{sinr} of these channels. 
        If user~$j$ increases the transmit power,
        receivers of these channels (\gls{bs}~$k$) experience stronger interference.
        \item $\mathcal{C}^3(c_{ij}) = \{ c_{ik}, k = 1, \dots, N, k \neq j \}$. These are channels that contribute to the interference of \gls{bs}~$i$.
        Channel gains $g_{ik}$ appears in the denominator of the \gls{sinr} of channel $c_{ij}$.
        If transmitters of these channels (user~$k$) increase their transmit power,
        receiver of channel $c_{ij}$ (\gls{bs}~$i$) experiences stronger interference.
        \item $\mathcal{C}^4(c_{ij}) = \{ c_{kp}, k, p = 1, \dots, N, k \neq i, p \neq j \}$. These are all other channels that do not contribute directly to the performance of link from user~$j$ to \gls{bs}~$i$.
    \end{enumerate}
    These categories are illustrated in Figure~\ref{fig:categories}.
    \item The transmitter-receiver pairs in the interference channel are permutation-equivariant.
    As explained in Section~\ref{sec:equivariance}.
\end{itemize}

\begin{figure}
    \centering
    \resizebox{.3\textwidth}{!}{
    \begin{tikzpicture}[scale=1]
\draw (0,0) -- (4,0) -- (4,4) -- (0,4) -- (0,0);
\draw (0, 1.2) -- (4, 1.2);
\draw (0, 1.6) -- (4, 1.6);
\draw (0.7, 0) -- (0.7, 4);
\draw (1.1, 0) -- (1.1, 4);

\node at (0.9, 1.4) {1};
\node at (0.9, 0.6) {2};
\node at (0.9, 2.8) {2};
\node at (0.35, 1.4) {3};
\node at (2.55, 1.4) {3};
\node at (0.35, 0.6) {4};
\node at (0.35, 2.8) {4};
\node at (2.55, 2.8) {4};
\node at (2.55, 0.6) {4};
\node at (-0.3, 1.4) {$i$};
\node at (0.9, -0.3) {$j$};
\node at (2, -1) {User (transmitter)};
\node at (-1, 2) [rotate=90] {BS (receiver)};
    \end{tikzpicture}
    }
    \caption{Illustration of 4 categories of channels for the channel $c_{ij}$: Category~1 is $c_{ij}$ itself. Category~2 contains channels sharing the same user with $c_{ij}$ but with different BSs. Category~3 contains channels sharing the same BS with $c_{ij}$ but with different users. Category~4 are the remaining channels.}
    \label{fig:categories}
\end{figure}

Based on the above observations,
we design an \gls{nn} architecture \InterferenceNet for optimization problems with the interference channel in such a way,
that we apply the same information processing to all channels for each category defined above,
we organize the information flow according to the 4 categories 
by aggregating information within a category
and concatenating information of different categories,
and we preserve the permutation-equivariance.
Two building blocks of the \InterferenceNet are the categorization of channels and information aggregation within a category.
The former is based on our domain knowledge of communication
whereas the latter is inspired by other successful \gls{nn} techniques, such as average pooling~\cite{yu2014mixed} and 
message aggregation in \gls{gnn}~\cite{fey2019fast}.
\end{change}

\subsection{Network Architecture}
The \InterferenceNet has $L$ layers.
The channel from user~$j$ to \gls{bs}~$i$ is represented by a feature vector $\mathbf{f}_{ij, l}$ as the input of layer~$l$.
In particular, $\mathbf{f}_{ij, 1} = \log_{10}(g_{ij})$, 
where $g_{ij}$ is the channel gain as defined in \eqref{eq:problem}. 


For $l < L$, 
the output feature vector of channel~$c_{ij}$ in layer~$l$ (i.e., the input feature in layer~$l + 1$) in category~$d$ is computed as
\begin{equation}
\mathbf{f}^d_{ij, l + 1} = \sum_{c_{kp} \in \mathcal{C}^d(c_{ij})}\text{ReLU}(\mathbf{W}_l^d \mathbf{f}_{kp, l} + \mathbf{b}_l^d) / |\mathcal{C}^d(i, j)|,
\label{eq:nic_layer}
\end{equation}
where $\mathbf{W}^d_l$ and $\mathbf{b}^d_l$ are the trainable weights and bias of layer~$l$ for category~$d$, respectively,
\begin{change}
i.e., we apply the same information processing in style of a fully connected layer $\text{ReLU}(\mathbf{W}_l^d \mathbf{f}_{kp, l} + \mathbf{b}_l^d)$ to the feature vector of every channel $c_{kp}$ in category~$d$ of channel $c_{ij}$,
then compute the mean of the output feature vectors of channels in this category.
\end{change}
The reason we use the mean rather than the sum is that different categories have different numbers of channels.
For example, if we have 7 users,
category~1 has 1 channel and category~4 has 36 channels.
If we use the sum instead of the mean,
the variance of features in category~4 would be much higher than the variance of features in category~1.
This is numerically difficult for the neural network to learn.
Therefore,
we use the mean instead of the sum to make the features of different categories roughly identically distributed.

The total output feature vector of channel~$c_{ij}$ in layer~$l$ is the concatenation of channel gain and the feature vectors in all four categories, i.e.,
\begin{equation}
    \mathbf{f}_{ij, l + 1} = \left(\log_{10}(g_{ij}), \mathbf{f}^1_{ij, l + 1}, \mathbf{f}^2_{ij, l + 1}, \mathbf{f}^3_{ij, l + 1}, \mathbf{f}^4_{ij, l + 1}\right).
\end{equation}
\begin{change}
That is to say,
original channel gain and information from all four categories are concatenated for every channel
as the input of the next layer.
\end{change}
For $l < L$, $\mathbf{f}_{ij, l + 1}$ is the input for the next layer.

For $l = L$ (i.e., for the last layer), the output of channel $c_{jj}$ is computed as
\begin{equation}
    f_{jj, L + 1} = \mathbf{w}_L \mathbf{f}_{jj, L} + b_L,
\end{equation}
where the output $f_{jj, L + 1}$ is a scalar instead of a vector (unlike the previous layers), 
$\mathbf{w}_L$ and $b_L$ are trainable weights and bias of layer $L$, respectively.
Note that only the diagonal elements are processed in the last layer (i.e., only $c_{jj}$ is processed and $c_{ij}$ is omitted for $i \neq j$)
since the output per user-\gls{bs} pair is a scalar,
therefore the indices are $f_{jj, L + 1}$ instead of $f_{ij, L + 1}$,
which is the action of user~$j$.

The above calculation is easy for readers to understand but difficult to be parallelized.
In fact,
we can organize the input features of layer~$l$ of all channels as a three-dimensional tensor $\mathbf{F}_l$ of shape $(\text{dim}(\mathbf{f}_{ij, l}), I, I)$,
where  $\mathbf{F}_l[\cdot, i, j] = \mathbf{f}_{ij,l}$.
For $l < L$,
the output feature tensors in categories~$1$-$4$ are computed by
\begin{align}
    \mathbf{F}_{l + 1}^1 &= \relu(\mathbf{W}_l^1 \bullet \mathbf{F}_l + \mathbf{b}_l^1\bullet \mathbf{1}_{1 \times I \times I})
    \label{eq:c1_computation}\\
    \mathbf{F}_{l + 1}^2 &= \mathbf{E} \circ\relu(\mathbf{W}_l^2 \bullet \mathbf{F}_l + \mathbf{b}_l^2\bullet \mathbf{1}_{1 \times I \times I}) / (I - 1)
    \label{eq:c2_computation}\\
    \mathbf{F}_{l + 1}^3 &= \relu(\mathbf{W}_l^3 \bullet \mathbf{F}_l + \mathbf{b}_l^3\bullet \mathbf{1}_{1 \times I \times I})\circ\mathbf{E}  / (I - 1)
    \label{eq:c3_computation}\\
    \mathbf{F}_{l + 1}^4 &= \mathbf{E} \circ\relu(\mathbf{W}_l^4 \bullet \mathbf{F}_l + \mathbf{b}_l^4\bullet \mathbf{1}_{1 \times I \times I})\circ\mathbf{E}  / ((I - 1)^2)
    \label{eq:c4_computation}\,,
\end{align}
respectively.
The final output $\mathbf{F}_{l + 1}$ is the concatenation of $\log_{10}(\mathbf{G})$, $\mathbf{F}_{l + 1}^1$, $\mathbf{F}_{l + 1}^2$, $\mathbf{F}_{l + 1}^3$, and $\mathbf{F}_{l + 1}^4$ along the first dimension.
In this way, the computation in the neural network can be performed efficiently with tensor products.

In the last layer, the raw output is computed as
\begin{equation}
    \mathbf{f}_{L + 1} = \mathbf{w}_L\mathrm{diag}(\mathbf{F}_L) + b_L \mathbf{1}_{1 \times N}.
    \label{eq:final_computation}
\end{equation}

For $\alpha_\delta$, the final layer does not have an activation function.
For $\beta_o$, a special activation function is necessary
since a uniform distribution $\mathcal{U}(\mathbf{a}, \mathbf{a} + \ell)$ can only be defined when every dimension of $\ell$ is positive,
output of $\beta_o$ is defined as 
\begin{equation}
\ell = \max(\mathbf{f}_{L+1}^{\beta}, \ell_{\min}),
\label{eq:bound_l}
\end{equation}
where $\mathbf{f}_{L + 1}^{\beta}$ is the raw output of the last layer of $\beta_o$
and $\ell_{\min}$ is a very small number (e.g., $10^{-6}$).
For all other layers in both $\alpha_\beta$ and $\beta_o$,
the activation function is ReLU.

\begin{figure*}[htbp]
    \centering
    \subfigure[First layer]{
    \resizebox{.8\textwidth}{!}{
    \begin{tikzpicture}
y={(0.5cm,0.25cm)},x={(0.5cm,-0.25cm)},z={(0cm,{veclen(0.5,0.25)*1cm})}
    ]
    \DrawCubes [step=10mm,thin]{0}{6}{4}{5}{0}{6}
    \node (width) [rectangle, yshift=1cm, xshift=.5cm,font=\Large] {Transmitter};
    \node (height) [rectangle, yshift=2.1cm, xshift=-2.3cm, anchor=east,font=\Large] {Channel};
    \node (channels) [rectangle, yshift=2.8cm, xshift=5.5cm,rotate=45,font=\Large] {Receiver};

\node[draw, single arrow,
              minimum height=25mm, minimum width=30mm,
              single arrow head extend=2mm,
              anchor=west, rotate=-45] at (6.7, 1.5) {};

\node (layer1) [layer, xshift=12cm, yshift=1.2cm,font=\Large] {Layer $l$ for category 1};
\node (layer2) [layer, below of=layer1, yshift=0cm,font=\Large] {Layer $l$ for category 2};
\node (layer3) [layer, below of=layer2, yshift=0cm,font=\Large] {Layer $l$ for category 3};
\node (layer4) [layer, below of=layer3, yshift=0cm,font=\Large] {Layer $l$ for category 4};

\draw [->] (3.7, 2.2) -- (15.7, 2.2);
\draw [->] (layer1.east) -- (15.7, 1.2);
\draw [->] (layer2.east) -- (15.7, 0.2);
\draw [->] (layer3.east) -- (15.7, -0.8);
\draw [->] (layer4.east) -- (15.7, -1.8);

    \DrawCubes [step=10mm,thin]{17.99}{24}{0}{5}{0}{6}
    \node (width) [rectangle, yshift=-3cm, xshift=18.5cm,font=\Large] {Transmitter};
    \node (height) [rectangle, yshift=4.5cm, xshift=24cm, anchor=west,font=\Large] {Channel};
    \node (height) [rectangle, yshift=3.5cm, xshift=24cm, anchor=west,font=\Large] {Category 1};
    \node (height) [rectangle, yshift=2.5cm, xshift=24cm, anchor=west,font=\Large] {Category 2};
    \node (height) [rectangle, yshift=1.5cm, xshift=24cm, anchor=west,font=\Large] {Category 3};
    \node (height) [rectangle, yshift=0.5cm, xshift=24cm, anchor=west,font=\Large] {Category 4};
    \node (channels) [rectangle, xshift=14cm, yshift=-1.2cm, xshift=9.5cm,rotate=45,font=\Large] {Receiver};
\end{tikzpicture}
    }}
    \subfigure[Intermediate layers]{
    \resizebox{.8\textwidth}{!}{
    \begin{tikzpicture}
y={(0.5cm,0.25cm)},x={(0.5cm,-0.25cm)},z={(0cm,{veclen(0.5,0.25)*1cm})}
    ]
    \DrawCubes [step=10mm,thin]{0}{6}{0}{5}{0}{6}
    \node (width) [rectangle, yshift=-3cm, xshift=.5cm,font=\Large] {Transmitter};
    \node (height) [rectangle, yshift=2.1cm, xshift=-2.3cm, anchor=east,font=\Large] {Channel};
    \node (height) [rectangle, yshift=1.1cm, xshift=-2.3cm, anchor=east,font=\Large] {Category 1};
    \node (height) [rectangle, yshift=0.1cm, xshift=-2.3cm, anchor=east,font=\Large] {Category 2};
    \node (height) [rectangle, yshift=-0.9cm, xshift=-2.3cm, anchor=east,font=\Large] {Category 3};
    \node (height) [rectangle, yshift=-1.9cm, xshift=-2.3cm, anchor=east,font=\Large] {Category 4};
    \node (channels) [rectangle, yshift=-1.2cm, xshift=5.5cm,rotate=45,font=\Large] {Receiver};

\node[draw, single arrow,
              minimum height=25mm, minimum width=30mm,
              single arrow head extend=2mm,
              anchor=west] at (6.7, -0.5) {};

\node (layer1) [layer, xshift=12cm, yshift=1.2cm,font=\Large] {Layer $l$ for category 1};
\node (layer2) [layer, below of=layer1, yshift=0cm,font=\Large] {Layer $l$ for category 2};
\node (layer3) [layer, below of=layer2, yshift=0cm,font=\Large] {Layer $l$ for category 3};
\node (layer4) [layer, below of=layer3, yshift=0cm,font=\Large] {Layer $l$ for category 4};

\draw [->] (3.7, 2.2) -- (15.7, 2.2);
\draw [->] (layer1.east) -- (15.7, 1.2);
\draw [->] (layer2.east) -- (15.7, 0.2);
\draw [->] (layer3.east) -- (15.7, -0.8);
\draw [->] (layer4.east) -- (15.7, -1.8);

    \DrawCubes [step=10mm,thin]{17.99}{24}{0}{5}{0}{6}
    \node (width) [rectangle, yshift=-3cm, xshift=18.5cm,font=\Large] {Transmitter};
    \node (height) [rectangle, yshift=4.5cm, xshift=24cm, anchor=west,font=\Large] {Channel};
    \node (height) [rectangle, yshift=3.5cm, xshift=24cm, anchor=west,font=\Large] {Category 1};
    \node (height) [rectangle, yshift=2.5cm, xshift=24cm, anchor=west,font=\Large] {Category 2};
    \node (height) [rectangle, yshift=1.5cm, xshift=24cm, anchor=west,font=\Large] {Category 3};
    \node (height) [rectangle, yshift=0.5cm, xshift=24cm, anchor=west,font=\Large] {Category 4};
    \node (channels) [rectangle, xshift=14cm, yshift=-1.2cm, xshift=9.5cm,rotate=45,font=\Large] {Receiver};
\end{tikzpicture}
    }}
    \subfigure[Final layer]{
    \resizebox{.8\textwidth}{!}{
    \begin{tikzpicture}
y={(0.5cm,0.25cm)},x={(0.5cm,-0.25cm)},z={(0cm,{veclen(0.5,0.25)*1cm})}
    ]
    \DrawCubes [step=10mm,thin]{0}{6}{0}{5}{0}{6}
    \node (width) [rectangle, yshift=-3cm, xshift=.5cm,font=\Large] {Transmitter};
    \node (height) [rectangle, yshift=2.1cm, xshift=-2.3cm, anchor=east,font=\Large] {Channel};
    \node (height) [rectangle, yshift=1.1cm, xshift=-2.3cm, anchor=east,font=\Large] {Category 1};
    \node (height) [rectangle, yshift=0.1cm, xshift=-2.3cm, anchor=east,font=\Large] {Category 2};
    \node (height) [rectangle, yshift=-0.9cm, xshift=-2.3cm, anchor=east,font=\Large] {Category 3};
    \node (height) [rectangle, yshift=-1.9cm, xshift=-2.3cm, anchor=east,font=\Large] {Category 4};
    \node (channels) [rectangle, yshift=-1.2cm, xshift=5.5cm,rotate=45,font=\Large] {Receiver};

\node[draw, single arrow,
              minimum height=25mm, minimum width=30mm,
              single arrow head extend=2mm,
              anchor=west] at (7.7, 0.2) {};

\node (layer2) [layer, xshift=12cm, yshift=0.2cm, yshift=0cm,font=\Large] {Layer $L$};
\draw [->] (layer2.east) -- (15.7, 0.2);

    \DrawCubes [step=10mm,thin]{17.99}{24}{2}{3}{0}{6}
    \node (width) [rectangle, yshift=-1cm, xshift=18.5cm,font=\Large] {Transmitter};
    \node (height) [rectangle, yshift=2.5cm, xshift=24cm, anchor=west,font=\Large] {Transmit power};
    \node (channels) [rectangle, xshift=14cm, yshift=0.8cm, xshift=9.5cm,rotate=45,font=\Large] {Receiver};
    
    \begin{scope}[canvas is xz plane at y=\YGridMax]
    \draw [fill=black!20!white]  (17.99,0) rectangle (18.99,1);
    \draw [fill=black!20!white]  (18.99,1) rectangle (19.99,2);
    \draw [fill=black!20!white]  (19.99,2) rectangle (20.99,3);
    \draw [fill=black!20!white]  (20.99,3) rectangle (21.99,4);
    \draw [fill=black!20!white]  (21.99,4) rectangle (22.99,5);
    \draw [fill=black!20!white]  (22.99,5) rectangle (23.99,6);
    \end{scope}
    
    \begin{scope}[canvas is xy plane at z=\ZGridMax]
    \draw [fill=black!20!white]  (22.99,2) rectangle (23.99,3);
    \end{scope}
    
    \begin{scope}[canvas is yz plane at x=\XGridMax]
    \draw [fill=black!20!white]  (2,5) rectangle (3,6);
    \end{scope}
\end{tikzpicture}
    }}
    \caption{Illustration of information processing from first over intermediate to final output layers.}
    \label{fig:info_processing}
\end{figure*}
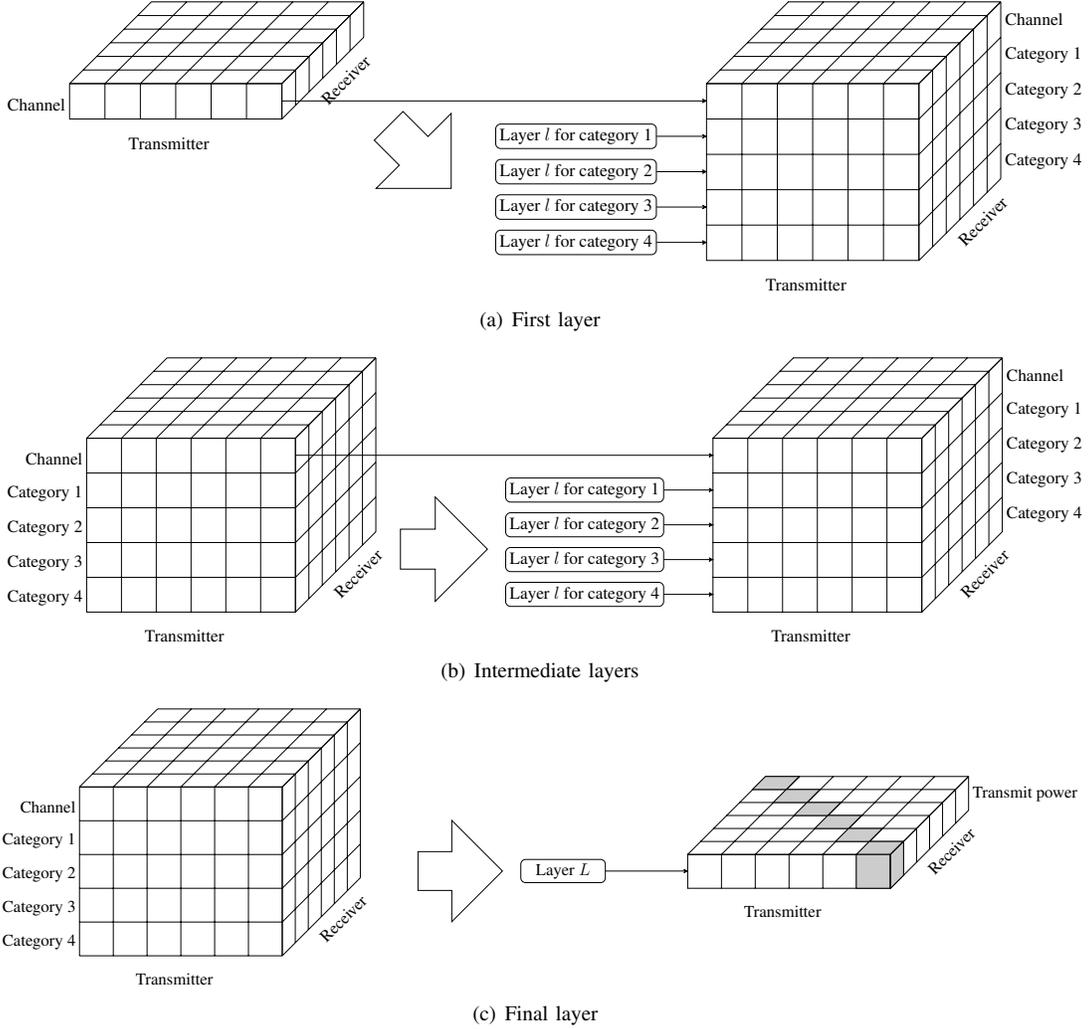

The above-described information processing is illustrated in Figure~\ref{fig:info_processing}.
An important advantage of the proposed method is the low complexity due to the small number of trainable parameters
because all channels use the same filters (\eqref{eq:c1_computation} - \eqref{eq:c4_computation}) to process the information.
Therefore, the number of trainable neural network parameters is independent from the number of user-\gls{bs} pairs.
In the proposed architecture,
there are 39844 trainable parameters.
Compared to it, the neural network with fully connected layers proposed in~\cite{matthiesen2020globally} has 8708189 trainable parameters,
which is approximately 218 times more than the \InterferenceNet architecture.

Besides the complexity,
another advantage of the proposed \InterferenceNet architecture is its permutation-equivariance.
Intuitively, it is because the above-defined 4 categories do not change by any permutation and 
the summation within a category is permutation-invariant.
Mathematically, the permutation-equivariance is proved as follows.

\begin{theorem}
An \InterferenceNet $\Phi$ maps from channel $\mathbf{G}$ to power allocation $\mathbf{p}$,
i.e., $\mathbf{p} = \Phi(\mathbf{G})$.
For any permutation matrix $\mathbf{P}$, 
we have $\mathbf{p} \mathbf{P}^T = \Phi(\mathbf{P} \mathbf{G} \mathbf{P}^T)$.
\end{theorem}

\begin{proof}
See Appendix~\ref{sec:proof_theorem1}.
\end{proof}

\subsection{Equivalence to Graph Neural Network and Distributed Implementation}

\begin{figure}[htbp]
    \centering
    \resizebox{.25\textwidth}{!}{
    \begin{tikzpicture}[
roundnode/.style={circle, draw=black, minimum size=7mm},
]
\node[roundnode] (node1) {$\mathbf{f}_{i,l}$};
\node[roundnode] (node2) [above of=node1, xshift=2cm] {$\mathbf{f}_{j,l}$};
\node[roundnode] (node3) [left of=node2, xshift=-2.5cm, yshift=0cm] {$\mathbf{f}_{k,l}$};

\draw[->] (node2) -- node [above] {$\mathbf{e}_{j, i}$} (node1);
\draw[->] (node3) -- (node1);
\end{tikzpicture}
    }
    \caption{A simple graph on which a GNN is defined.}
    \label{fig:gnn}
\end{figure}
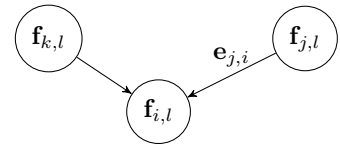

The \gls{gnn} is a novel neural network architecture that performs information processing for multiple nodes connected by (directional) edges~\cite{johnson2016composing}.
A simple example of a \gls{gnn} is shown in Figure~\ref{fig:gnn}.

The node feature of node $i$ on layer $l$ is computed as~\cite{fey2019fast}
\begin{equation}
    \mathbf{f}_{i,l} = \gamma_{l} \left( \mathbf{f}_{i,l-1}, \square_{j \in \mathcal{N}(i)} \phi_{l}\left( \mathbf{f}_{i,l-1}, \mathbf{f}_{j,l-1}, \mathbf{e}_{j, i} \right) \right),
    \label{eq:gnn}
\end{equation}
where $\mathbf{f}_{i,l}$ is the node feature of node~$i$ on layer~$l$,
$\mathbf{e}_{j,i}$ is the edge feature of the directional edge from node~$j$ to node~$i$,
$\phi_{l}$ is a neural network that is applied on each edge and performs message passing 
(in the example in \eqref{eq:gnn}, the edge is from node~$j$ to node~$i$, therefore its input is the concatenation of $\mathbf{f}_{i,l-1}$, $\mathbf{e}_{j,i}$ and $\mathbf{f}_{j,l-1}$),
$\square$ is a symmetric, permutation-invariant function (i.e., the function value is constant if the arguments are permuted, for example, $\square(x_1, x_2, \dots, x_n) = \square(x_2, x_1, \dots, x_n)$), such as sum or max,
$\gamma_{l}$ is a neural network that is applied to each node on layer~$l$ and processes the information of the node itself and the messages from its neighbors.
In the example in \eqref{eq:gnn}, the node is node~$i$, therefore its input is the concatenation of $\mathbf{f}_{i,l-1}$ and $\square_{j \in \mathcal{N}(i)} \phi_{l}\left( \mathbf{f}_{i,l-1}, \mathbf{f}_{j,l-1}, \mathbf{e}_{j, i} \right)$.
The distributed nature of the \gls{gnn} and its ability to perform message passing and information aggregation make it a competitive candidate to optimize cellular networks~\cite{eisen2020optimal,wang2022learning,chowdhury2021unfolding}.

Although the \InterferenceNet architecture is introduced as a centralized neural network,
we can understand each row in Figure~\ref{fig:categories} as a node and
easily prove that it can be reshaped to fit into the form of \eqref{eq:gnn}.
Based on \eqref{eq:c1_computation} - \eqref{eq:c4_computation},
we define $\phi$, $\square$ and $\gamma$ as
\begin{equation}
\begin{aligned}
& \phi_{l}(\mathbf{F}_l[\cdot, i, \cdot])\\
=&\left(
\begin{array}{l}
    \text{ReLU}(\mathbf{W}_l^2 \bullet \mathbf{F}_l[\cdot, i, \cdot] + \mathbf{b}_l^2\bullet \mathbf{1}_{1 \times I \times I})\\
    \text{ReLU}(\mathbf{W}_l^4 \bullet \mathbf{F}_l[\cdot, i, \cdot] + \mathbf{b}_l^4\bullet \mathbf{1}_{1 \times I \times I})\circ\mathbf{E}  / (I - 1)
\end{array}
\right),
\end{aligned}
    \label{eq:phi_gnn}
\end{equation}

\begin{equation}
    \square_{j\in \mathcal{N}(i)} \phi_{l}(\mathbf{F}_l[\cdot, j, \cdot]) = \sum_{j}\phi_{l}(\mathbf{F}_l[\cdot, j, \cdot]) / (I - 1),
    \label{eq:square_gnn}
\end{equation}
and
\begin{equation}
\begin{aligned}
& \gamma_{l}(\mathbf{F}_l[\cdot, i, \cdot], \square_{j\in \mathcal{N}(i)} \phi_{l}(\mathbf{F}_l[\cdot, j, \cdot]))\\
= & \left(
\begin{array}{l}
    \mathbf{G}\\
    \text{ReLU}(\mathbf{W}_l^1 \bullet \mathbf{F}_l[\cdot, i, \cdot] + \mathbf{b}_l^1\bullet \mathbf{1}_{1 \times I \times I})\\
    \mathbf{E}\circ\text{ReLU}(\mathbf{W}_l^3 \bullet \mathbf{F}_l[\cdot, i, \cdot] + \mathbf{b}_l^3\bullet \mathbf{1}_{1 \times I \times I})  / (I - 1)\\
    \square_{j\in \mathcal{N}(i)} \phi_{l}(\mathbf{F}_l[\cdot, j, \cdot])
\end{array}
\right),
\end{aligned}
\label{eq:gamma_gnn}
\end{equation}
respectively,
where the concatenation in \eqref{eq:phi_gnn} and \eqref{eq:gamma_gnn} is along the first dimension,
we conclude that the computation in one layer is presented in the form of \eqref{eq:gnn}.
This fact proves that the proposed \InterferenceNet architecture is equivalent to the \gls{gnn},
thus inherits its advantages such as distributed implementation in each \gls{bs} and limited data amount of message passing in the fronthaul.

Summarizing the above sections,
the proposed algorithm to solve the non-convex power control problem is formulated as Algorithm~\ref{alg:alg}.

\begin{algorithm}
\caption{Neural network training with a non-convex objective with stochastic action and reparameterization trick}\label{alg:alg}
\begin{algorithmic}
\State Formulate the objective function according to \eqref{eq:objective2}.
\State Initialize $\alpha_\delta$ and $\beta_o$ such that $\alpha_\delta(\mathbf{G}) < 0$ and $\alpha_\delta(\mathbf{G}) + \beta_o(\mathbf{G}) > p_{\max}$ for all possible $\mathbf{G}$.
\State Initialize $\kappa = 0$.
\While{$H(\alpha_\delta(\mathbf{G}), \beta_o(\mathbf{G})) > H_0$}
\State Perform a gradient ascent step of \eqref{eq:objective2} w.r.t. $\delta$ and $o$.
\State Update $\kappa$ according to \eqref{eq:updating_kappa}.
\EndWhile
\end{algorithmic}
\end{algorithm}
\section{Training and Testing Results}
\label{sec:results}

In this section, we evaluate the introduced \InterferenceNet architecture on a multi-user communication scenario.

\subsection{Scenario and Channel Models}
\label{sec:scenario}

We consider a multi-cell mobile network scenario with four \glspl{bs}, as shown in Figure~\ref{fig:scenario}.
Coordinates of the four \glspl{bs} are $(0.5, 0.5)$, $(0.5, 1.5)$, $(1.5, 1.5)$ and $(0.5, 0.5)$, with the unit of $\si{\km}$,
and each \gls{bs} is equipped with $n_R=2$ antennas.
\begin{change}
Each user is assigned to the closest \gls{bs} and
the \glspl{bs} use matched filter to process the received signal.
\end{change}
We consider $I=7$ and $I=4$ single-antenna users who share the same \gls{prb} and thus have interference on each other.
Please note that the total number of users in this cellular network can be more than $I$ because a cellular network can have more than one \gls{prb}.
Users allocated with different \glspl{prb} have negligible interference on each other because their signals are well separated in time or frequency domain.
Furthermore, the transmit power control problems of different \glspl{prb} can be considered decoupled from each other and can be solved separately.
The network can therefore serve many more users while we focus on the difficult part:
users sharing the same \gls{prb} and causing interference to each other.
The same power control strategy can be applied to every \gls{prb} individually.

Assumptions on channel models and parameters are in line with \cite{matthiesen2020globally} in general.
In particular,
the propagation path-loss is computed with the model presented in~\cite{calcev2007wideband}.
The carrier frequency is assumed to be \SI{1.8}{\GHz} whereas the power decay factor is \num{4.5}.
The fast fading is modeled as \gls{iid} complex random variables,
where the amplitude is drawn from the standard normal distribution
and the complex phase is uniformly distributed between $0$ and $2\pi$.
The static power consumption is assumed to be $\SI{1}{\watt}$
and the power amplifier inefficiency $\mu_i=4$ for all $i$.
Noise power at the receiver is computed as $\sigma^2=FN_0B$,
where $F=\SI{3}{\decibel}$ is the noise figure at the receiver,
$N_0=\SI{-174}{\dBm\per\Hz}$ is the noise spectrum density,
$B=\SI{180}{\kilo\hertz}$ is the bandwidth.
All users have the same maximum transmit power $p_{\max}$.

The \InterferenceNet is implemented with PyTorch 1.11,
in which the reparameterization trick is implemented as the \emph{rsample} method.
There are $L=5$ layers in the \InterferenceNet
and the feature dimension is 20 for each category and each layer.
The ADAM optimizer is chosen to train the neural network
and the learning rate is set to be $2\cdot 10^{-7}$ for $p_{\max} > \SI{0}{\dBm}$ and $10^{-7}$ otherwise.
The training set contains 6000 samples whereas the testing set has 1000 samples.
The batch size is 512.

\subsection{Training and Testing Results}

\begin{figure}
    \centering
    \input{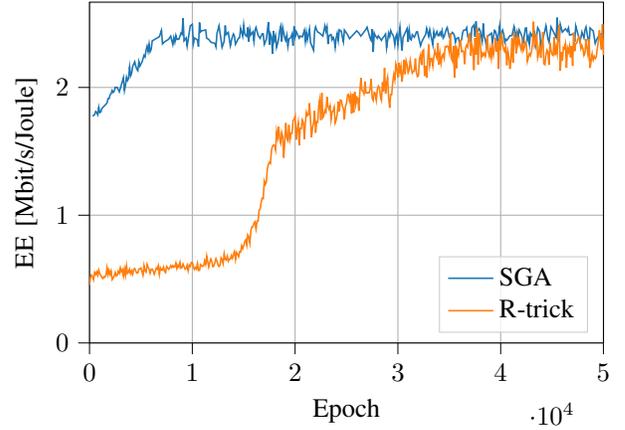}
    \caption{Improvement of objective with training with 4 users using SGA and reparameterization trick.}
    \label{fig:training_4users}
\end{figure}

As mentioned above, we consider 4 and 7 users.
Figure~\ref{fig:training_4users} shows the improvement of objective using \gls{sga} and reparameterization trick with 4 users.
The maximum transmit power is \SI{5}{dBm}.
Although the \gls{ee} optimization problem is non-convex,
\gls{sga} is able to obtain the same \gls{ee} as the reparameterization trick,
indicating that both methods find the same solution.
In the following text,
we will therefore focus on the results with 7 users.

\begin{figure}
    \centering 
    \input{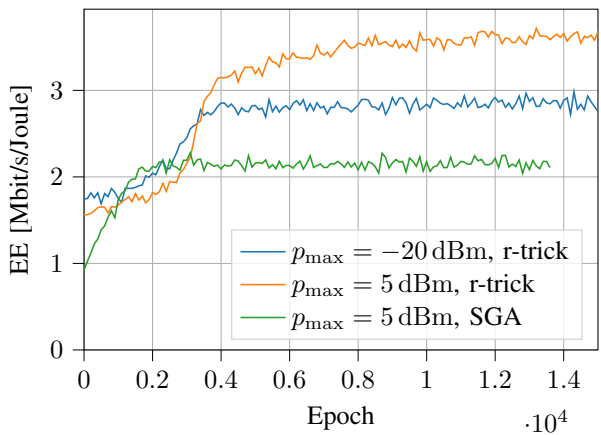}
    \caption{Improvement of objective with training over 15000 thousand epochs for three cases with 7 users.}
    \label{fig:training}
\end{figure}

Figure~\ref{fig:training} shows the curves of achieved \gls{ee} with 7 users during training.
The blue, orange and green curves are achieved \gls{ee} with $p_{\max} = \SI{-20}{\dBm}$ and reparameterization trick,
$p_{\max} = \SI{5}{\dBm}$ and reparameterization trick,
and $p_{\max} = \SI{5}{\dBm}$ and conventional \gls{sga}, respectively.
All three setups use the same neural network architecture and hyperparameters except the learning rate (see Section~\ref{sec:scenario}).
When the maximum transmit power is low ($\SI{-20}{\dBm}$),
the optimal transmit power is the maximum transmit power $p_{\max}$ and the achieved \gls{ee} is limited by it.
When the maximum transmit power is high ($\SI{5}{\dBm}$),
the optimal transmit power lies between 0 and $p_{\max}$
and the achieved \gls{ee} is higher than with a low maximum transmit power.
Unlike the four-user scenario,
if we use the conventional \gls{sga} instead of the objective function \eqref{eq:objective2},
it converges to a poor local optimum much earlier than training with the reparameterization trick,
as the green curve shows.
This result shows the advantage of the proposed solution compared to the conventional \gls{sga} for non-convex objectives.

\begin{figure}
    \centering
    \input{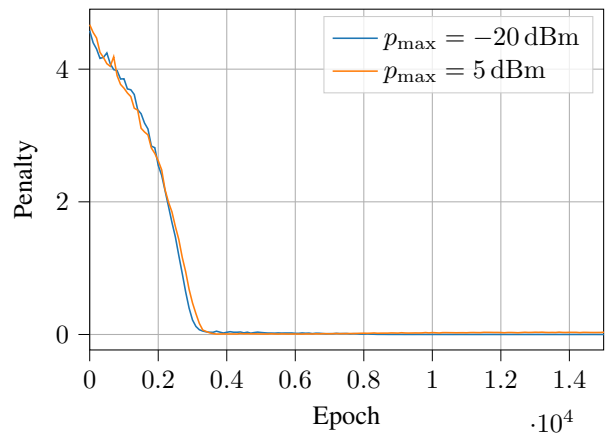}
    \caption{Value of penalty during training.}
    \label{fig:penalty}
\end{figure}

In order to gain more insights into the training process,
Figure~\ref{fig:penalty} shows the penalty introduced by \eqref{eq:penalty1} and \eqref{eq:penalty2}.
We can observe that the penalty reduces to 0 quickly at the beginning of the training,
suggesting that $\mathbf{a}$ and $\mathbf{b}$ lies within $[0, p_{\max}]$ for all $\mathbf{G}$ after training of some epoches.
Please note that it is important to initialize $\mathbf{a}$ and $\mathbf{b}$ outside the feasible region because the global optimum cannot be considered in \eqref{eq:objective2} if it lies outside $[\mathbf{a}, \mathbf{b}]$
(an example is, as shown above, the optimal transmit power is $p_{\max}$ for a low maximum transmit power).
Since it is very difficult to initialize $\mathbf{a}$ and $\mathbf{b}$ exactly on the boundaries,
it is easier to initialize them outside the feasible region and use the penalty terms to reduce them to the feasible region at the beginning of the training.

\begin{figure}
    \centering
    \input{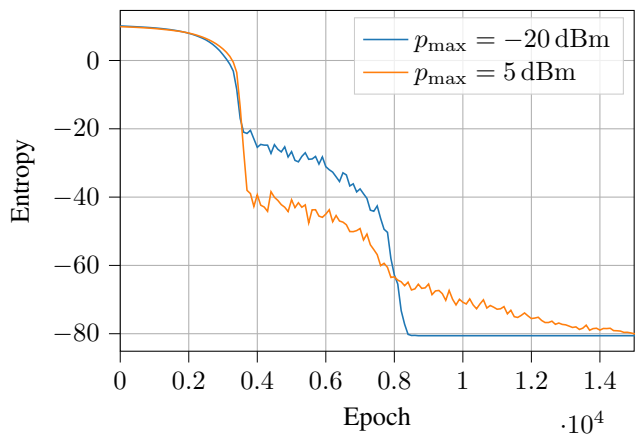}
    \caption{Value of entropy of action distribution during training.}
    \label{fig:entropy}
\end{figure}

Figure~\ref{fig:entropy} illustrates the mean entropy during training for the two aforementioned maximum transmit powers.
Since $\mathbf{a}$ and $\ell$ are initialized in such a way that the initial transmit power distribution covers the whole interval $[0, p_{\max}]$ for any $\mathbf{G}$.
As a result, the entropy is high at the beginning of the training.
It begins to decrease quickly first due to the penalty terms \eqref{eq:penalty1} and \eqref{eq:penalty2}
because the penalty term is independent from $\mathbf{G}$.
Since $p_{\max}$ is assumed to be the same for all users,
the penalty term is the same for all samples.
This results in the high decreasing speed of entropy.
After $\mathbf{a}$ and $\mathbf{a} + \ell$ lie within $[0, p_{\max}]$,
the entropy decreases more slowly because the objective function is different given different $\mathbf{G}$.
The entropy reduces to roughly \num{-80} at the end of training due to the lower bound of $\ell$ posed in \eqref{eq:bound_l}.
Training with higher maximum transmit power converges more slowly due to the larger and more complex feasible region.

\begin{figure}
    \centering
    \input{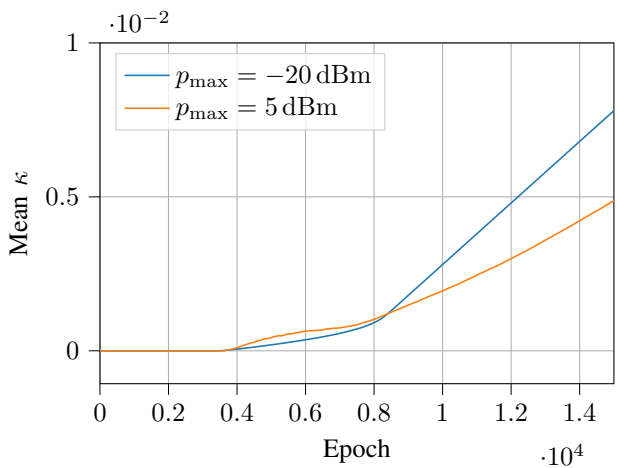}
    \caption{Value of $\kappa$ during training.}
    \label{fig:kappa}
\end{figure}

Figure~\ref{fig:kappa} shows the mean $\kappa$ during training.
As described in Section~\ref{sec:entropy},
$\kappa$ is initialized to be 0 
and is increased if the entropy does not decrease in the previous $h$~epochs.
We can observe that the mean $\kappa$ is 0 at first
and begins to increase after about \num{4000} epochs,
indicating that the entropy of the distribution does not decrease for some samples (as shown in Figure~\ref{fig:nonconvex_objective-below-local}).
Comparing Figure~\ref{fig:entropy} and Figure~\ref{fig:kappa},
we can see that the mean entropy is far above the final entropy as $\kappa$ begins to increase (after about 4000 epochs),
indicating that the increasing $\kappa$ is due to multiple local optima rather than the lower bound of the distribution \eqref{eq:bound_l}.
An increased $\kappa$ helps to overcome the suboptimal local optima and to further decrease the entropy.
Due to the lower bound on $\ell$ \eqref{eq:bound_l},
the entropy stops decreasing despite the increasing $\kappa$ in the final phase of training.
This happens after roughly \num{8000} epochs for $p_{\max}=\SI{-20}{\dBm}$ and \num{15000} epochs for $p_{\max}=\SI{5}{\dBm}$.
The increasing speed of mean $\kappa$ is slower for $p_{\max}=\SI{5}{\dBm}$ because the convergence speed with the high maximum transmit power is slower (see Figure~\ref{fig:entropy}).

\subsection{Comparison to Global Optimum and Baseline}
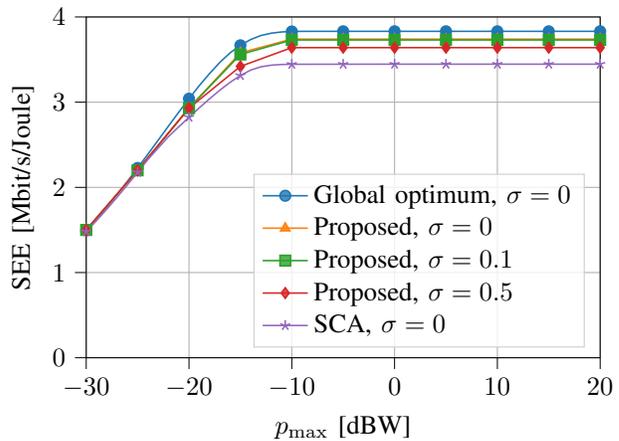
\begin{figure}
    \centering
\begin{tikzpicture}

\definecolor{crimson2143940}{RGB}{214,39,40}
\definecolor{darkgray176}{RGB}{176,176,176}
\definecolor{darkorange25512714}{RGB}{255,127,14}
\definecolor{forestgreen4416044}{RGB}{44,160,44}
\definecolor{lightgray204}{RGB}{204,204,204}
\definecolor{mediumpurple148103189}{RGB}{148,103,189}
\definecolor{sienna1408675}{RGB}{140,86,75}
\definecolor{steelblue31119180}{RGB}{31,119,180}

\begin{axis}[
legend cell align={left},
legend style={
  fill opacity=0.8,
  draw opacity=1,
  text opacity=1,
  at={(0.97,0.03)},
  anchor=south east,
  draw=lightgray204
},
tick align=outside,
tick pos=left,
x grid style={darkgray176},
xlabel={\(\displaystyle p_{\max}\) [dBW]},
xmajorgrids,
xmin=-30, xmax=20,
xtick style={color=black},
y grid style={darkgray176},
ylabel={SEE [Mbit/s/Joule]},
ymajorgrids,
ymin=0, ymax=4,
ytick style={color=black},
width=.95\linewidth,
height=.25\textheight,
]
\addplot [semithick, steelblue31119180, mark=*, mark repeat=5]
table {%
-30 1.4961674
-29 1.6293612
-28 1.7695304
-27 1.9165856
-26 2.0696225
-25 2.2272296
-24 2.389164
-23 2.554087
-22 2.7187726
-21 2.881994
-20 3.0406055
-19 3.193864
-18 3.3358982
-17 3.4670243
-16 3.578989
-15 3.667392
-14 3.735151
-13 3.7826304
-12 3.809573
-11 3.8240595
-10 3.829481
-9 3.8312294
-8 3.831646
-7 3.8316543
-6 3.8316567
-5 3.8316567
-4 3.8316567
-3 3.8316567
-2 3.8316567
-1 3.8316567
0 3.8316567
1 3.8316567
2 3.8316567
3 3.8316567
4 3.8316567
5 3.8316567
6 3.8316567
7 3.8316567
8 3.8316567
9 3.8316567
10 3.8316567
11 3.8316567
12 3.8316567
13 3.8316567
14 3.8316567
15 3.8316567
16 3.8316567
17 3.8316567
18 3.8316567
19 3.8316567
20 3.8316567
};
\addlegendentry{Global optimum, $\sigma=0$}
\addplot [semithick, darkorange25512714, mark=triangle*, mark repeat=1]
table {%
-30 1.5
-25 2.2
-20 2.93
-15 3.58
-10 3.74
-5 3.74
0 3.74
5 3.74
10 3.74
15 3.74
20 3.74
};
\addlegendentry{Proposed, $\sigma=0$}
\addplot [semithick, forestgreen4416044, mark=square*, mark repeat=1]
table {%
-30 1.5
-25 2.2
-20 2.93
-15 3.56
-10 3.73
-5 3.73
0 3.73
5 3.73
10 3.73
15 3.73
20 3.73
};
\addlegendentry{Proposed, $\sigma=0.1$}
\addplot [semithick, crimson2143940, mark=diamond*, mark repeat=1]
table {%
-30 1.5
-25 2.2
-20 2.93
-15 3.42
-10 3.64
-5 3.64
0 3.64
5 3.64
10 3.64
15 3.64
20 3.64
};
\addlegendentry{Proposed, $\sigma=0.5$}
\addplot [semithick, mediumpurple148103189, mark=star, mark repeat=5]
table {%
-30 1.4811945
-29 1.6096524
-28 1.7441728
-27 1.8843174
-26 2.0288694
-25 2.1768284
-24 2.3275404
-23 2.46156028
-22 2.59558016
-21 2.71048936
-20 2.82349731
-19 2.93344245
-18 3.03944575
-17 3.14065516
-16 3.23297035
-15 3.31313756
-14 3.37630515
-13 3.41544381
-12 3.43417794
-11 3.44220251
-10 3.44462803
-9 3.44500754
-8 3.44506538
-7 3.44511743
-6 3.4452148
-5 3.44536758
-4 3.4455634
-3 3.44572732
-2 3.44584522
-1 3.44590558
0 3.44592916
1 3.44593618
2 3.44593908
3 3.44593947
4 3.44594
5 3.44594
6 3.44594
7 3.44594
8 3.44594
9 3.44594
10 3.44594
11 3.44594
12 3.44594
13 3.44594
14 3.44594
15 3.44594
16 3.44594
17 3.44594
18 3.44594
19 3.44594
20 3.44594
};
\addlegendentry{SCA, $\sigma=0$}
\end{axis}

\end{tikzpicture}
    \caption{Comparison between global optimum, achieved performance and baseline algorithms.}
    \label{fig:global_results}
\end{figure}

\begin{change}
Above text shows details and unveils insights of the training.
In Figure~\ref{fig:global_results},
we show the achieved \gls{ee} with the testing data
(i.e., data that have not been used in the training phase)
and compare it to the global optimum computed with the branch-and-bound algorithm
and the \gls{sca} algorithm with double-initialization approach
for 7 users and different maximum transmit powers (details of the global optimum and baselines are available in~\cite{matthiesen2020globally}).
\end{change}
The optimal transmit power is the maximum transmit power when it is low (therefore the global optimum is easier to find)
and it lies between $[0, p_{\max}]$ for higher maximum transmit power
(therefore the \gls{ee} stops increasing with further increasing $p_{\max}$),
which makes the problem more challenging.
\begin{change}
From the figure we observe that the proposed method outperforms the baseline algorithm \gls{sca},
and achieves good results close to the global optimum.
Furthermore, if we assume imperfect \gls{csi} by adding Gaussian perturbations with certain standard deviation to the \gls{nn} input,
the achieved \glspl{ee} with standard deviations $\sigma=0.1$ and $\sigma=0.5$ still outperform the baseline.
This result proves that the \InterferenceNet does not require perfect \gls{csi} to generate the near-optimal power allocation.
\end{change}

\begin{change}
Since the curves in Figure~\ref{fig:global_results} are close to each other and the difference between them is not easy to recognize,
Table~\ref{tab:diff} lists the relative differences of the proposed approach and the \gls{sca} algorithm compared to the global optimum for $p_{\max}=\SI{0}{dBW}$ (absolute difference to the global optimum divided by the global optimum).
We can see that the proposed method realizes an \gls{ee} very close to the global optimum and considerably better than the \gls{sca} algorithm.
Besides, the performance is robust to the \gls{csi} perturbation.
\begin{table}[htbp]
    \centering
    \caption{Relative difference of each approach compared to global optimum}
    \label{tab:diff}
    \begin{tabular}{ll}
        \toprule
        Approach & Relative difference\\
        \midrule
        Proposed & 2.39\% \\
        Proposed ($\sigma=0.1$) & 2.65\% \\
        Proposed ($\sigma=0.5$) & 5.01\% \\
        SCA & 10.07\% \\
        \bottomrule
    \end{tabular}
\end{table}
\end{change}

\subsection{Resilience Against Setting Mismatch}
\begin{change}
In order to verify the resilience of setting mismatch of the trained model,
we apply the model trained with 7 users to test data with 4 users,
which is technically possible due to the property that the number of \InterferenceNet parameters is independent from the number of user-\gls{bs} pairs.
However, this mismatch is obviously more severe than possible mismatches in operation (e.g., unconventional channel gains due to extreme user positions).
The result is listed in Table~\ref{tab:generality}.
We can see that the model trained with 7 users has slightly worse performance than model trained with 4 users (no setting mismatch in training and testing) and the \gls{sca} algorithm.
On the contrary, the model trained with 4 users cannot generate a reasonable \gls{ee} with 7 users.
This result illustrates the resilience against setting mismatch and the \enquote{downward compatibility} of the proposed solution.

\begin{table}[htbp]
    \centering
    \caption{Achieved EE with different methods for generality check of the trained model.}
    \label{tab:generality}
    \begin{tabular}{ll}
        \toprule
        Algorithm & Achieved EE with 4 users [Mbit/s/Joule]\\
        \midrule
        Global optimum & 2.48\\
        Model trained with 4 users & 2.44\\
        Model trained with 7 users & 2.38\\
        SCA & 2.43\\
        \bottomrule
    \end{tabular}
\end{table}
\end{change}

\subsection{Complexity}
\label{sec:complexity}
\begin{change}
Since the \InterferenceNet only has multiplication, summation and ReLU operation,
the complexity of the neural network is linear to the number of user-\gls{bs} pairs.
Compared to that, the branch-and-bound algorithm has exponential worst-case complexity in general.
The \gls{sca} algorithm has a polynomial complexity for a convex surrogate function.
From a more practical point of view,
information processing of each user-\gls{bs} pair in one layer is independent from each other and can therefore be parallized on GPU or TPU.
While the branch-and-bound algorithm takes several minutes 
and the \gls{sca} algorithm needs multiple seconds to compute the power allocation for one data sample,
the computation of power allocation with \InterferenceNet is done in 7 milliseconds on a Macbook Pro with the M2 Pro processor.
This result shows that the proposed method not only achieves a high performance,
but also has a low operational complexity.
\end{change}

\section{Conclusion}
\label{sec:conclusion}

The \gls{ee} is an important objective of the future wireless communication systems.
Due to the non-convexity of the objective,
this problem has multiple local optima and cannot be solved with conventional convex optimization tools.
Although the global optimum can be found with the branch-and-bound algorithm,
this solution does not scale well due to the high complexity.
\begin{change}
This paper presents an unsupervised machine learning solution to the problem,
which optimizes the function (realized by an \gls{nn}) that maps from the channel matrix to the transmit power with respect to the objective via \gls{sga}.
Two fundamental building blocks of unsupervised learning are objective function and \gls{nn} architecture.
Main contribution of this work includes an objective function for non-convex problems and an innovative \gls{nn} architecture \InterferenceNet.
\end{change}
We first introduce a reparameterization trick based objective.
Instead of optimizing one operation point of transmit powers,
we apply stochastic action,
let the transmit powers uniformly distributed in the interval $[\mathbf{a}, \mathbf{b}]$
and define the objective as the expected \gls{ee} given the transmit power distribution,
where we approximate the expectation with the mean of random variables and
use the reparameterization trick to make the random variable differentiable.
The second contribution is a dedicated \gls{nn} architecture \emph{\InterferenceNet},
which encodes the domain knowledge of interference channels and realizes permutation-equivariance (i.e., when the \gls{bs}-user pairs are permuted,
the output of the neural network is permuted in the same way automatically).
The \InterferenceNet defines four categories of channels for each user-\gls{bs} pair according to their impact on the \gls{sinr}.
Training and testing results show solid improvement behavior.
Compared to the state-of-the-art,
the achieved \gls{ee} is close to the global optimum obtained with the brand-and-bound algorithm
and is better than the \gls{sca} algorithm.
\begin{change}
Besides the \gls{ee} optimization,
methods proposed in this work can also be applied to other purposes:
On one hand, we can apply the proposed objective function to other non-convex problems,
since the reparameterization-based objective function is universal.
On the other hand, we can apply the \InterferenceNet to different optimization problems related to \gls{sinr} in interference channels,
such as sum-rate optimization,
because \InterferenceNet is designed for the interference channel and is not necessarily assosiated to \gls{ee}.
In addition,
we can expand this work by, e.g., considering minimum rate requirement of each user.
\end{change}
Source code and data are available under \url{https://github.com/bilepeng/ee}.

\section*{Acknowledgement}

The authors would like to thank Mr. F. Siegismund-Poschmann for his suggestion on the test function and Mr. R. Wang for helping the training and testing process,
as well as the editor and anonymous reviewers for helping us improving the manuscript quality.


\appendices
\section{Proof of Theorem 1}
\label{sec:proof_theorem1}

We first extend the permutation of matrix \eqref{eq:permutation_matrix} to permutation of tensor as
\begin{equation}
    \mathbf{F}' = \mathbf{P} \circ \mathbf{F} \circ \mathbf{P}^T,
    \label{eq:permutation}
\end{equation}
i.e., the second and third dimensions (corresponding to the rows and columns of a matrix)
are permuted by $\mathbf{P}$ simultaneously.
Then we prove that the operation of layer~$l$ is permutation-equivairant, i.e.,
    $\mathbf{P} \circ \mathbf{F}_{l + 1}\circ \mathbf{P}^T = \Phi_l(\mathbf{P}\circ \mathbf{F}_l\circ \mathbf{P}^T)$
holds if $\mathbf{F}_{l + 1} = \Phi_l(\mathbf{F}_l)$ for $l < L$,
where $\Phi_l$ is the information processing in layer~$l$, and
    $\mathbf{p} \mathbf{P}^T = \Phi_L(\mathbf{P}\circ \mathbf{F}_L\circ \mathbf{P}^T)$
holds if $\mathbf{p} = \Phi_L(\mathbf{F}_L)$.
Finally, we prove the whole \InterferenceNet is also equivariant.

Combining \eqref{eq:c1_computation} and \eqref{eq:permutation}, the permuted output for category~1 is computed as
\begin{equation}
    \begin{aligned}
    (\mathbf{F}_{l + 1}^1)' = & \text{ReLU}(\mathbf{W}_l^1 \bullet \mathbf{P} \circ \mathbf{F}_l \circ \mathbf{P}^T + \mathbf{b}_l^1\bullet \mathbf{1}_{1 \times N \times N})\\
    = & \text{ReLU}(\mathbf{P} \circ \mathbf{W}_l^1 \bullet \mathbf{F}_l \circ \mathbf{P}^T\\ 
    & + \mathbf{P} \circ\mathbf{b}_l^1\bullet \mathbf{1}_{1 \times N \times N}\circ \mathbf{P}^T)\\
    = & \mathbf{P} \circ\text{ReLU}( \mathbf{W}_l^1 \bullet \mathbf{F}_l + \mathbf{b}_l^1\bullet \mathbf{1}_{1 \times N \times N})\circ \mathbf{P}^T\\
    = & \mathbf{P} \circ \mathbf{F}_{l + 1}^1 \circ \mathbf{P}^T.
    \end{aligned}
    \label{eq:c1_permutation}
\end{equation}
In the second line of \eqref{eq:c1_permutation},
exchanging order of $\mathbf{P}$ and $\mathbf{W}_l^1$ is allowed because the multiplication between $\mathbf{W}_l^1$ and $\mathbf{F}_l$ is in the first dimension
and does not affect the permutation in the second and third dimension.
Multiplying $\mathbf{P}$ and $\mathbf{P}^T$ on the left and right hand sides of $\mathbf{b}_l^1\bullet \mathbf{1}_{1 \times I \times I})$ in the second and third dimensions
does not change $\mathbf{b}_l^1\bullet \mathbf{1}_{1 \times I \times I}$ because this action permutes its second and third dimension,
which are identical.
In the third line of \eqref{eq:c1_permutation},
we move the permutation out of the summation and ReLU operation because the ReLU is an elementwise operation and does not depend on the order of elements in the tensor.
Therefore, the operation of a layer on category~1 is permutation-equivariant.

In order to prove the permutation-equivariance of the other three categories,
we first prove that the product between $\mathbf{E}$ and any permutation matrix $\mathbf{P}$ is commutative,
which is straightforward:
\begin{equation}
    \begin{aligned}
    \mathbf{E}\mathbf{P} & = \mathbf{1} \mathbf{P} - \mathbf{I} \mathbf{P}
     = \mathbf{P} \mathbf{1} - \mathbf{P} \mathbf{I}
     = \mathbf{P} \mathbf{E},
    \end{aligned}
    \label{eq:commutative}
\end{equation}
where we use the property $\mathbf{1} \mathbf{P} = \mathbf{P} \mathbf{1} = \mathbf{1}$ because the sum of rows and columns of $\mathbf{P}$ is 1.

Combining \eqref{eq:c2_computation} and \eqref{eq:permutation}, the permuted output for category~2 is computed as
\begin{equation}
    \begin{aligned}
    (\mathbf{F}_{l + 1}^2)' = & \mathbf{E} \circ\text{ReLU}(\mathbf{W}_l^2 \bullet \mathbf{P} \circ \mathbf{F}_l \circ \mathbf{P}^T \\
    & + \mathbf{b}_l^2\bullet \mathbf{1}_{1 \times N \times N}) / (2(N - 1))\\
    = & \mathbf{E} \circ\mathbf{P} \circ\text{ReLU}( \mathbf{W}_l^2 \bullet \mathbf{F}_l \\
    & + \mathbf{b}_l^2\bullet \mathbf{1}_{1 \times N \times N})\circ \mathbf{P}^T / (2(N - 1))\\
    = & \mathbf{P} \circ\mathbf{E} \circ\text{ReLU}( \mathbf{W}_l^2 \bullet \mathbf{F}_l \\
    & + \mathbf{b}_l^2\bullet \mathbf{1}_{1 \times N \times N})\circ \mathbf{P}^T / (2(N - 1))\\
    = & \mathbf{P} \circ \mathbf{F}_{l + 1}^2 \circ \mathbf{P}^T.
    \end{aligned}
    \label{eq:c2_permutation}
\end{equation}
In the second line of \eqref{eq:c2_permutation},
we reuse the result of \eqref{eq:c1_permutation}.
In the third line of \eqref{eq:c2_permutation},
we use the fact proved above that the product between $\mathbf{E}$ and $\mathbf{P}$ is commutative.
Proofs of equivariance for categories~3 and 4 are similar to proof for category~2 and are omitted.
Since the operation of a layer is equivariant for all four categories,
the operation is equivariant for the total feature as well,
i.e., $\mathbf{P} \mathbf{F}_{l + 1} \mathbf{P}^T = \Phi_l(\mathbf{P} \mathbf{F}_l \mathbf{P}^T)$ holds
if $\mathbf{F}_{l + 1} = \Phi_l( \mathbf{F}_l)$ for $l < L$.

For the last layer, the proof of equivariance is similar to the previous layers because the additional diagonalization operation is also equivariant.

Having proved all layers in \InterferenceNet is equivariant,
it is easy to prove that the \InterferenceNet itself is also equivariant by recursively pulling $\mathbf{P}$ out of the operators:
\begin{equation}
    \begin{aligned}
    \mathbf{p}' = & \Phi_L(\dots\Phi_2(\Phi_1(\mathbf{P} \circ\mathbf{G}\circ \mathbf{P}^T))\dots)\\
    = & \Phi_L(\dots\Phi_2(\mathbf{P} \circ\Phi_1(\mathbf{G})\circ \mathbf{P}^T)\dots)\\
    = & \dots \\
    = & \Phi_L(\Phi_{L - 1}(\dots \Phi_1(\mathbf{G})\dots) )\mathbf{P}^T\\
    = & \mathbf{p} \mathbf{P}^T.
    \end{aligned}
    \label{eq:equivariance_last_step}
\end{equation}
This concludes the proof.



\printbibliography

\begin{IEEEbiography}[{\includegraphics[trim={0 5cm 0 0},width=1in,height=1.25in,clip,keepaspectratio]{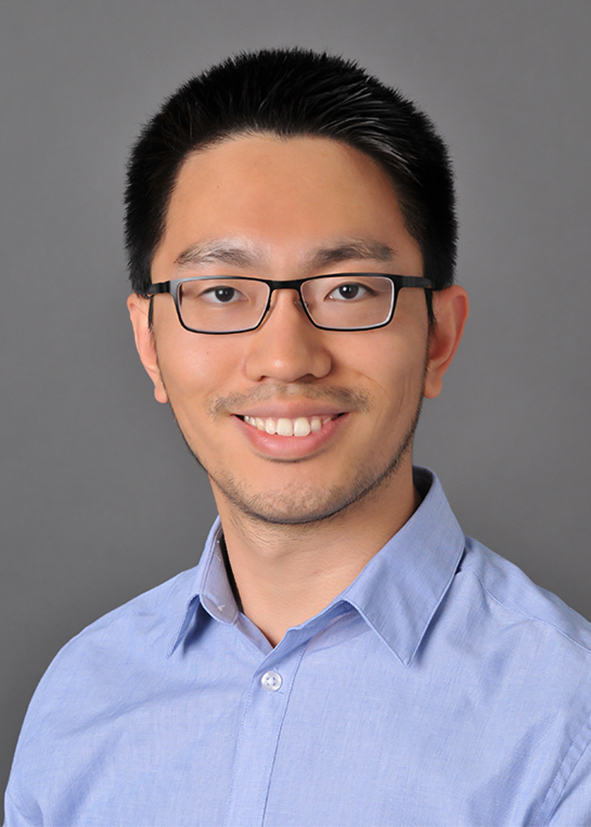}}]{Bile Peng}
 (Member, IEEE) received the Ph.D. degree with distinction from the Institute of Communications Technology, Technische Universit\"at Braunschweig in 2018. He has been a Postdoctoral researcher in the Chalmers University of Technology, Sweden from 2018 to 2019, a development engineer at IAV GmbH, Germany from 2019 to 2020. Currently, he is a Postdoctoral researcher in Institute of Communications Technology, Technische Universit\"at Braunschweig, Germany. His research interests include Bayesian inference and machine learning algorithms for signal processing and resource allocation of wireless communication systems.
He received the IEEE vehicular technology society 2019 Neal Shepherd memorial best propagation paper award.
\end{IEEEbiography}

\begin{IEEEbiography}[{\includegraphics[width=1in,height=1.25in,clip,keepaspectratio]{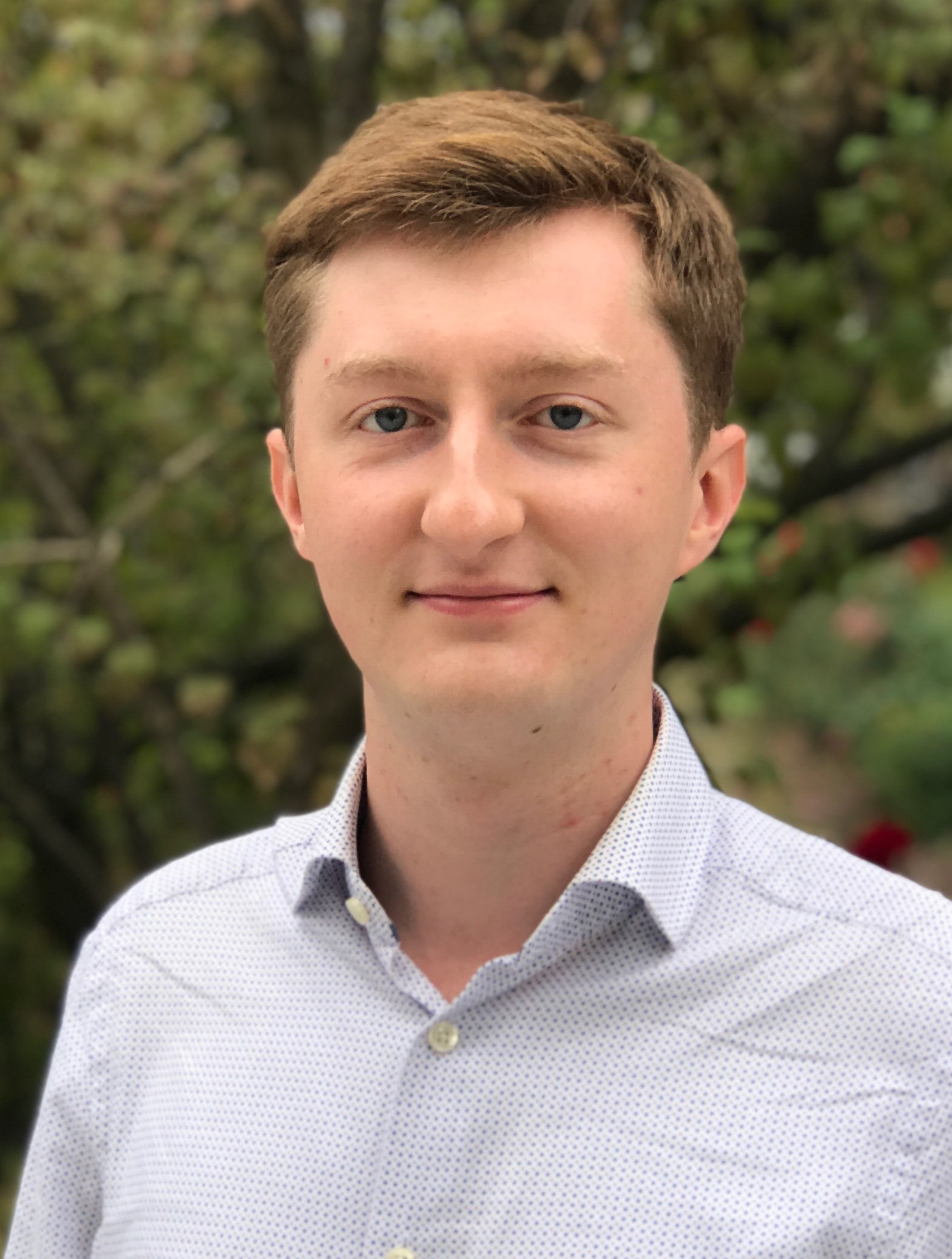}}]{Karl-Ludwig Besser} (Member, IEEE)
received his Dipl.-Ing. degree in electrical engineering from Technische Universit\"at Dresden, Germany in 2018 and his PhD in electrical engineering from Technische Universität Braunschweig, Germany in 2022.
In August 2018, he joined the Communications Theory group at TU Dresden.
From August 2019 until February 2023, he was with the Institute for Communications Technology at Technische Universit\"at Braunschweig, Germany.
Since March 2023, he is with the Department of Electrical and Computer Engineering at Princeton University.
His research interests include ultra-reliable communications, physical layer security, and the application of machine learning in communications.
\end{IEEEbiography}

\begin{IEEEbiography}[{\includegraphics[width=1in,height=1.25in,clip,keepaspectratio]{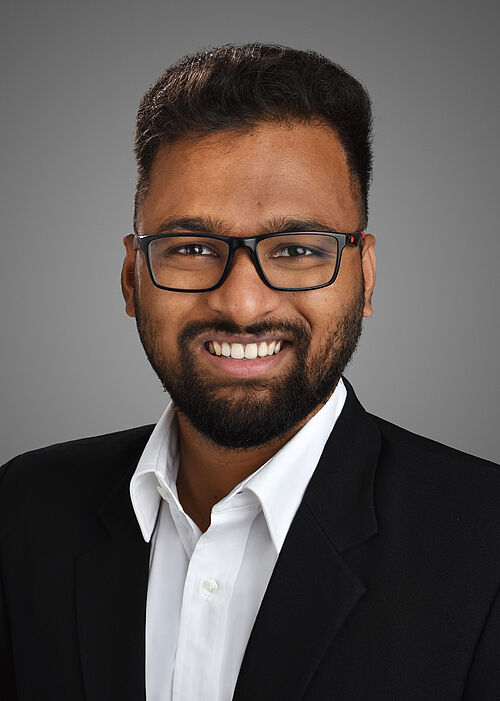}}]{Ramprasad Raghunath} (Student Member, IEEE)
has  graduated in the year 2021 from his master studies in Technische Universit\"at Braunschweig in the field of computational sciences in engineering with a specialization in mathematics and computer science.
He is currently a doctoral student in Institute of Communications Technology,
Technische Universit\"at Braunschweig.
His research area is machine learning algorithms for resource allocation problems in wireless communications.
\end{IEEEbiography}

\begin{IEEEbiography}
    [{\includegraphics[width=1in,height=1.25in,clip,keepaspectratio]{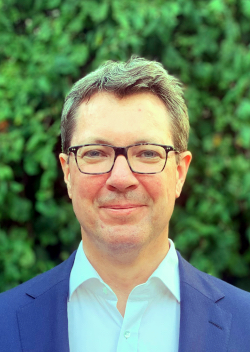}}]{Eduard A. Jorswieck} (Fellow, IEEE) is currently the Managing Director of the Institute of Communications Technology and the Head of the Chair for Communications Systems and a Full Professor at the Technische Universit\"at Braunschweig, Brunswick, Germany. From 2008 until 2019, he held the Chair for Communication Theory at TU Dresden. From 2006 until 2008, he was with the signal processing group at KTH Stockholm as post-doctoral fellow and assistant professor. Eduard has obtained his PhD in electrical engineering and computer science from TU Berlin in 2004. He has published more than 160 journal articles, 15 book chapters, one book, three monographs, and some 300 conference papers. His main research interests are in the broad area of communications. He was a recipient of the IEEE Signal Processing Society Best Paper Award. He and his colleagues were also recipients of the Best Paper and Best Student Paper Awards at the IEEE CAMSAP 2011, IEEE WCSP 2012, IEEE SPAWC 2012, IEEE ICUFN 2018, PETS 2019, and ISWCS 2019. Since 2017, he has been the Editor-in-Chief of the EURASIP Journal on Wireless Communications and Networking. Since 2022, he is on the editorial board of the IEEE TRANSACTIONS ON COMMUNICATIONS. He was on the editorial boards of the IEEE SIGNAL PROCESSING LETTERS, the IEEE TRANSACTIONS ON SIGNAL PROCESSING, the IEEE TRANSACTIONS ON WIRELESS COMMUNICATIONS, and the IEEE TRANSACTIONS ON INFORMATION FORENSICS AND SECURITY.
\end{IEEEbiography}

\end{document}